\documentclass[journal]{IEEEtran}

\ifCLASSINFOpdf
\else
\fi

\usepackage{amsmath,amsfonts,amssymb}
\usepackage{algorithmic}
\usepackage{array}

\usepackage{graphicx}
\usepackage{dcolumn}
\usepackage{bm}
\usepackage{hyperref}
\hypersetup{
	colorlinks   = true, 
	urlcolor     = blue, 
	linkcolor    = blue, 
	citecolor    = blue 
}
\usepackage{pifont}
\usepackage{multirow}
\usepackage{tabularray}
\usepackage{romannum}
\usepackage{bigints}
\usepackage{ulem}
\usepackage{xcolor}

\begin{document}

\title{Computational Electromagnetics Meets Spin Qubits: Controlling Noise Effects in Quantum Sensing and Computing}


\author{Wenbo~Sun,~\IEEEmembership{Graduate Student Member,~IEEE,}
        Sathwik Bharadwaj,~\IEEEmembership{Member,~IEEE,}
        Runwei Zhou,~\IEEEmembership{Graduate Student Member,~IEEE,}
        Dan~Jiao,~\IEEEmembership{Fellow,~IEEE,}
        and~Zubin~Jacob
\thanks{Wenbo Sun, Sathwik Bharadwaj, Runwei Zhou, Dan Jiao, and Zubin Jacob are with the Elmore Family School of Electrical and Computer Engineering, Purdue University, West Lafayette, IN, 47906 USA. Corresponding e-mail: zjacob@purdue.edu}
\thanks{Manuscript received xx, xx, 2024.}}

\markboth{Journal of \LaTeX\ Class Files,~Vol.~x, No.~x, xxxx~2024}%
{Shell \MakeLowercase{\textit{et al.}}: Bare Demo of IEEEtran.cls for IEEE Journals}

\maketitle

\begin{abstract}
Solid-state spin qubits have emerged as promising platforms for quantum information. Despite extensive efforts in controlling noise in spin qubit quantum applications, one important but less controlled noise source is near-field electromagnetic fluctuations. Low-frequency (MHz and GHz) electromagnetic fluctuations are significantly enhanced near lossy material components in quantum applications, including metallic/superconducting gates necessary for controlling spin qubits in quantum computing devices and materials/nanostructures to be probed in quantum sensing. Although controlling this low-frequency electromagnetic fluctuation noise is crucial for improving the performance of quantum devices, current efforts are hindered by computational challenges. In this paper, we leverage advanced computational electromagnetics techniques, especially fast and accurate volume integral equation based solvers, to overcome the computational obstacle. We introduce a quantum computational electromagnetics framework to control low-frequency magnetic fluctuation noise and enhance spin qubit device performance. Our framework extends the application of computational electromagnetics to spin qubit quantum devices. Furthermore, we demonstrate the application of our framework in realistic quantum devices. Our work paves the way for device engineering to control magnetic fluctuations and improve the performance of spin qubit quantum sensing and computing.
\end{abstract}


%
\IEEEpeerreviewmaketitle


\section{Introduction}
Solid-state spin qubits have emerged as promising quantum information platforms due to their small sizes, high controllability, and long preservation of encoded quantum information~\cite{burkard2023semiconductor}. In quantum applications based on spin qubits, quantum information is encoded in the spin degree of freedom of electrons or nuclei. Despite their advantages in sensing and information processing applications, quantum systems are generally susceptible to environmental noise~\cite{suter2016colloquium}.

To improve the performance of quantum applications, extensive efforts have been devoted to device engineering and controlling noise in spin qubit systems. For example, nuclear spins in the substrate can interact with spin qubits through hyperfine interactions and limit the performance of quantum computing and sensing devices~\cite{chekhovich2013nuclear}. Advanced material fabrication technologies, such as isotope engineering, have been developed to control the concentration of isotopes with non-zero nuclear spins and mitigate this noise~\cite{itoh2014isotope}. Fluctuating charges in semiconductor spin qubit systems can create random electric fields at qubit positions and perturb energy levels of spin qubits~\cite{kuhlmann2013charge,yoneda2023noise,boter2020spatial}. Recent devices employ undoped semiconductors and thin quantum wells to reduce charge noise and improve the quality of semiconductor spin qubits~\cite{burkard2023semiconductor,paquelet2023reducing,connors2019low}.

Meanwhile, another important yet less controlled noise in spin qubit applications originates from the near-field vacuum and thermal fluctuations of electromagnetic (EM) fields~\cite{langsjoen2012qubit,sun2023limits,degen2017quantum,kolkowitz2015probing,tenberg2019electron}. Spin qubits are less susceptible to noise electric fields, but are sensitive to magnetic field fluctuations in the environment. One common feature in spin qubit quantum applications is the extreme proximity of spin qubits to lossy materials. In quantum computing devices, nanofabricated metallic or superconducting gates and antennas are necessary for controlling semiconductor spin qubits~\cite{takeda2021quantum,huang2019fidelity,he2019two,morello2020donor,wang2016highly,xue2022quantum,hendrickx2021four}. In hybrid quantum systems, nanomagnets are employed to realize long-range control of spin qubits~\cite{awschalom2021quantum,niknam2022quantum}. In the near-field of these materials, evanescent waves associated with intrinsic material loss can significantly enhance the fluctuations of low-frequency ($\leq$GHz) magnetic fields~\cite{sun2024nano,khosravi2024giant,langsjoen2012qubit}. At GHz frequencies, the enhancement can exceed 15 orders of magnitude compared to free space~\cite{sun2024nano}. These significantly enhanced low-frequency magnetic fluctuations can interact with spin qubits, leading to noise effects that are important in spin qubit quantum applications. In quantum computing devices, these low-frequency magnetic fluctuations can accelerate the loss of quantum information and limit the devices' performance~\cite{sun2023limits,premakumar2017evanescent}. This motivates the necessity to model and suppress the low-frequency magnetic fluctuations in spin qubit quantum computing devices to improve their performance.

Meanwhile, in quantum sensing, shallow spin qubits are usually in the near-field of materials for probing material properties and imaging nanostructures~\cite{machado2023quantum,dolgirev2023local,kolkowitz2015probing,ariyaratne2018nanoscale,van2015nanometre,dwyer2022probing,staudacher2015probing}. Here, the near-field magnetic noise carries important information about the properties of the material system to be probed by spin qubits. Therefore, amplifying the low-frequency magnetic fluctuations can benefit the accuracy of quantum sensing of material properties. Furthermore, material systems of interest can be inhomogeneous (e.g., superconductors~\cite{bhattacharyya2024imaging} and two-dimensional magnets~\cite{thiel2019probing}) or nanostructured. Hence, accurate modeling of low-frequency magnetic fluctuations near arbitrarily structured lossy materials is also important for quantum sensing based on noise effects (e.g., quantum-impurity relaxometry~\cite{rovny2024new} or dephasometry).

Despite its importance, controlling the electromagnetic fluctuation noise remains less explored. Engineering the electromagnetic fluctuation noise can provide a new avenue in improving the performance of spin qubit quantum devices. However, current progress in this direction is largely limited by computational challenges.

Previous efforts in engineering electromagnetic fluctuations mainly focused on photonic environments, where high-frequency fluctuations are dominant in light-matter interactions. Computational methods based on differential equations, including finite-element methods (FEM) and finite-difference time-domain methods (FDTD), are used to model the photonic environments for engineering molecular spontaneous emission~\cite{baranov2017modifying}, atom-atom interactions~\cite{boddeti2021long,cortes2022fundamental}, near-field radiative heat transfer~\cite{otey2014fluctuational,rodriguez2011frequency}, and Casimir effects~\cite{reid2009efficient}, which are usually related to high-frequency ($\geq$THz) electromagnetic fluctuations. Recent work discussed full-wave FEM simulations of GHz electric noise for macroscopic superconducting transmon qubits~\cite{roth2021full,pham2023spectral}, which are typically larger than $100\,\mathrm{\mu m}$~\cite{wang2022towards}.

In contrast, controlling electromagnetic environments for spin qubits opens a new frontier in engineering electromagnetic fluctuation effects, where low-frequency (MHz and GHz) magnetic fluctuations in the near-field of ultra-subwavelength nanostructures become important~\cite{sun2024nano}. Here, electromagnetic fluctuations of interest are of MHz and GHz frequencies corresponding to wavelength $>1$cm, while lossy material components in spin qubit quantum devices can have a characteristic size of $\sim10$nm. Furthermore, due to evanescent wave contributions, this low-frequency magnetic fluctuation noise can have strong spatial variations in the nanometer range depending on its positions from nanostructures. Therefore, controlling magnetic fluctuation noise for spin qubits quantum applications requires accurately modeling the low-frequency ($\leq$GHz) electromagnetic environment near ultra-subwavelength nanoscale objects, which can lead to large memory and computation time consumption and highly ill-conditioned numerical system for FEM or FDTD methods.

\begin{figure}[!t]
    \centering
    \includegraphics[width=3in]{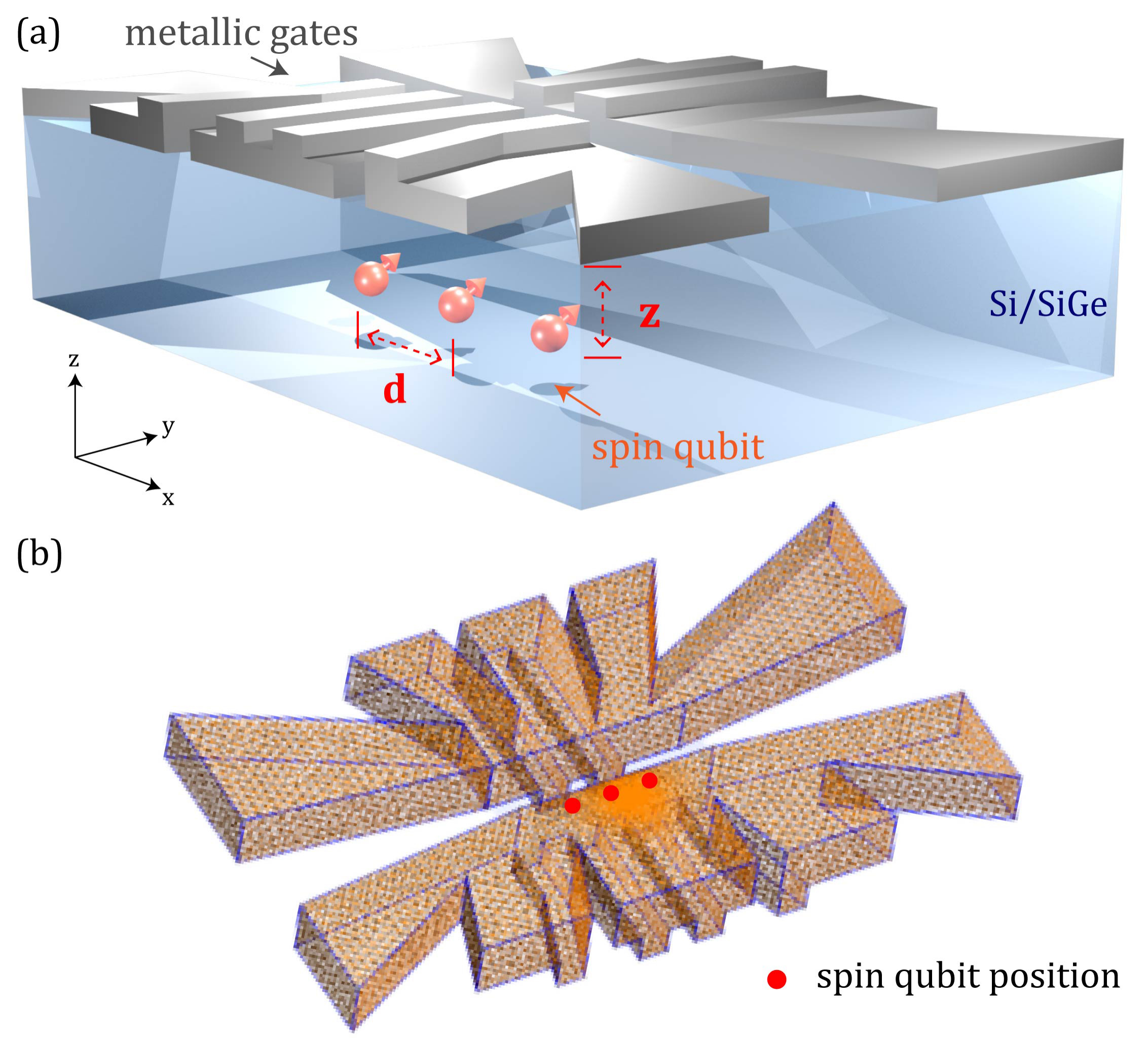}
    \caption{(a) A schematic of the simplified structure for a quantum computing device~\cite{takeda2021quantum} containing three semiconductor spin qubits. The spin qubits have interqubit distance $d$ and are at a distance $z$ from the top aluminum gates (grey structures) necessary for controlling spin qubits. (b) Tetrahedral mesh elements used in computational electromagnetic simulations to discretize the metal gates and solve volume integral equations. We increase the mesh density in the region close to spin qubits to improve the accuracy of computational electromagnetics simulations.}
    \label{fig:figure1}
\end{figure}

In this paper, we introduce a numerical framework to engineer low-frequency magnetic noise for enhancing spin qubit quantum sensing and computing performance by leveraging advanced computational electromagnetics techniques, especially fast and accurate volume integral equation (VIE) based solvers~\cite{2018_Maiomiao_Direct,omar2015linear,YifanWang_TAP2022}. Previous developments in modeling quantum devices with computational electromagnetics focused on superconducting qubits~\cite{roth2021full,tosi2023antenna,pham2023spectral,solgun2014blackbox,smith2016quantization}, cavity quantum electrodynamics (QED) systems~\cite{ryu2023matrix}, and two-photon interference~\cite{na2021diagonalization}. Here, we extend the application of computational electromagnetics to model noise effects in spin qubit quantum devices. With foundations established in our previous works~\cite{sun2023limits,sun2024nano}, the framework here takes advantage of volume integral equation methods to calculate comprehensive noise effects on spin qubits, including relaxation, dephasing, and open quantum system dynamics. We further apply the theoretical framework to control low-frequency magnetic noise in the nano-electromagnetic environment in realistic applications, including a three-semiconductor-spin-qubit quantum computing device (see Fig.~\ref{fig:figure1}) and nanoscale imaging based on quantum-impurity relaxometry (see Fig.~\ref{fig:figure5}). Additionally, through comparison with noise modeled by volume integral equations, we demonstrate the limitation of current approximate methods used in modeling magnetic noise in spin qubit systems, which can fail to predict qualitatively correct noise behaviors in realistic devices. Our work paves the way for controlling magnetic noise in spin qubit sensing and computing devices.

The paper is organized as follows. In Section~\ref{section2}, we review the fundamental working principles of solid-state spin qubits and important metrics for quantifying the noise effects on spin qubits. In Section~\ref{section3}, we discuss the origin of low-frequency electromagnetic fluctuations in solid-state spin qubit systems. We highlight that the obstacle to controlling magnetic fluctuation noise is accurately modeling the low-frequency magnetic fluctuations in the near field of nanostructured lossy materials. In Section~\ref{section4}, to overcome this obstacle, we introduce a quantum computational electromagnetics framework to leverage advanced computational electromagnetics techniques for modeling noise effects in spin qubit quantum applications. In Section~\ref{section_cem}, we demonstrate the application of fast and accurate volume integral equation based solvers in modeling low-frequency magnetic noise. In Section~\ref{section_qc}, we apply our theoretical framework to control low-frequency magnetic noise in a semiconductor spin qubit quantum computing device. In Section~\ref{section_qs}, we apply our theoretical framework for nanoscale imaging based on quantum-impurity relaxometry. Finally, in Section~\ref{section_conclusion}, we summarize the paper and indicate further applications of our framework in optimizing the design of quantum sensing and computing devices.

\section{Solid State Spin Qubit Preliminaries}\label{section2}
In this section, we review the fundamentals of solid-state-spin qubit systems. The progress of quantum computing platforms can provide a major edge in solving problems that are currently intractable with classical computing systems. The most promising near-term applications of quantum computers include simulations of microscopic properties of quantum materials \cite{daley2022practical}, high-energy physics \cite{bauer2023quantum}, and quantum chemistry \cite{bauer2020quantum}. This would have a significant impact on a variety of real-world applications, including catalysis \cite{von2021quantum} and the development of battery materials \cite{delgado2022simulating}. The basic building block of a qubit is a two-level system. The initialization, manipulation, and readout of the qubit states through optical, electronic, or microwave methods is at the core of the operation of any quantum circuits. Along with quantum computing applications, one can also leverage the sensitivity of qubits to their surroundings to develop quantum sensors \cite{degen2017quantum}. Despite their advantages in sensing and information processing applications, quantum systems are generally susceptible to environmental noise~\cite{suter2016colloquium}. Hence, the device design and packaging strategies \cite{rosenberg2020solid} based on computational electromagnetics are crucial as they enable a systematic assessment of the interaction with the device's electromagnetic environment.

\subsection{Introduction to Solid State Spin Qubits}
Solid-state spin qubits have emerged as leading platforms for implementing these promising quantum technology solutions \cite{ burkard2023semiconductor, chatterjee2021semiconductor}. Compact sizes, high controllability, long preservation time of quantum information, and proven semiconductor fabrication techniques \cite{zwerver2022qubits,neyens2024probing,sun2024full} offer an advantage for realizing scalable solid-state spin qubit quantum information platforms. In spin qubit systems, the two-level system is defined by the spin states of nuclear or electron spins. 
Various types of solid-state qubits, such as the Loss-DiVincenzo qubits \cite{loss1998quantum}, donor spin qubits \cite{kane1998silicon}, single-triplet spin qubits \cite{petta2005coherent}, exchange-only qubits \cite{bacon2000universal}, and vacancy/defect center qubits \cite{miao2020universal}, have been experimentally demonstrated.  

\subsubsection{Semiconductor Spin Qubits}
Spin qubits have been realized in semiconductor systems using four primary approaches. Loss-DiVincenzo qubits encode and decode information by altering the intrinsic spins of electrons trapped inside quantum dots~\cite{loss1998quantum}. Donor spin qubits (e.g., phosphorous or erbium) use the electron or nuclear spin of implanted donors in the substrate in place of quantum dots~\cite{morello2020donor}. Unlike Loss-DiVincenzo and donor spin qubits, two electrons are used in singlet-triplet spin qubits, and whether their spins are aligned or opposed is employed to encode information~\cite{petta2005coherent}. In exchange-only qubits, quantum information is encoded in the total spin angular momentum subspace of three or more electron spins, and gate operations are performed via the exchange interactions \cite{bacon2000universal}. In most spin qubit architectures, metallic/superconducting gates and antennas are required for qubit operation (Fig.~\ref{fig:figure1}), initialization, and readout, necessitating device design insights from computational electromagnetics.

\subsubsection{Vacancy centers}
Shallow vacancy center spin qubits are widely used in quantum sensing applications. Here, spin qubits are constructed based on vacancy centers in the substrate material. One well-known example is the negatively charged nitrogen-vacancy (NV) center spin qubit, which is constituted of a nitrogen atom near a vacancy site in the diamond. The energy band diagram of the negatively charged NV center involves a triplet ground state, a triplet excited state, and two intermediate singlet states \cite{gali2019ab}. The ground state triplet consists of three spin sublevels $m_s = 0$ and $m_s = \pm 1$ separated by zero-field splitting around $2.87$GHz. These ground state sublevels are usually selected as the two-level system needed to form the spin qubit~\cite{childress2013diamond} since they have a long coherence time exceeding $1$ ms at room temperatures~\cite{herbschleb2019ultra}. Several other vacancy centers have been demonstrated recently, including tin-vacancy centers in diamond \cite{debroux2021quantum}, defect centers in silicon carbide \cite{anderson2022five}, and spin defects in hexagonal boron nitride monolayers \cite{vaidya2023quantum}. In many spin qubit quantum sensing applications, spin qubits need to be in the near-field of materials, which can be inhomogeneous or have complex geometries at the nanoscale. This indicates the necessity to accurately model the electromagnetic environment near inhomogeneous or nanostructured materials through computational electromagnetics.

\begin{figure}
    \centering
    \includegraphics[width=3.1in]{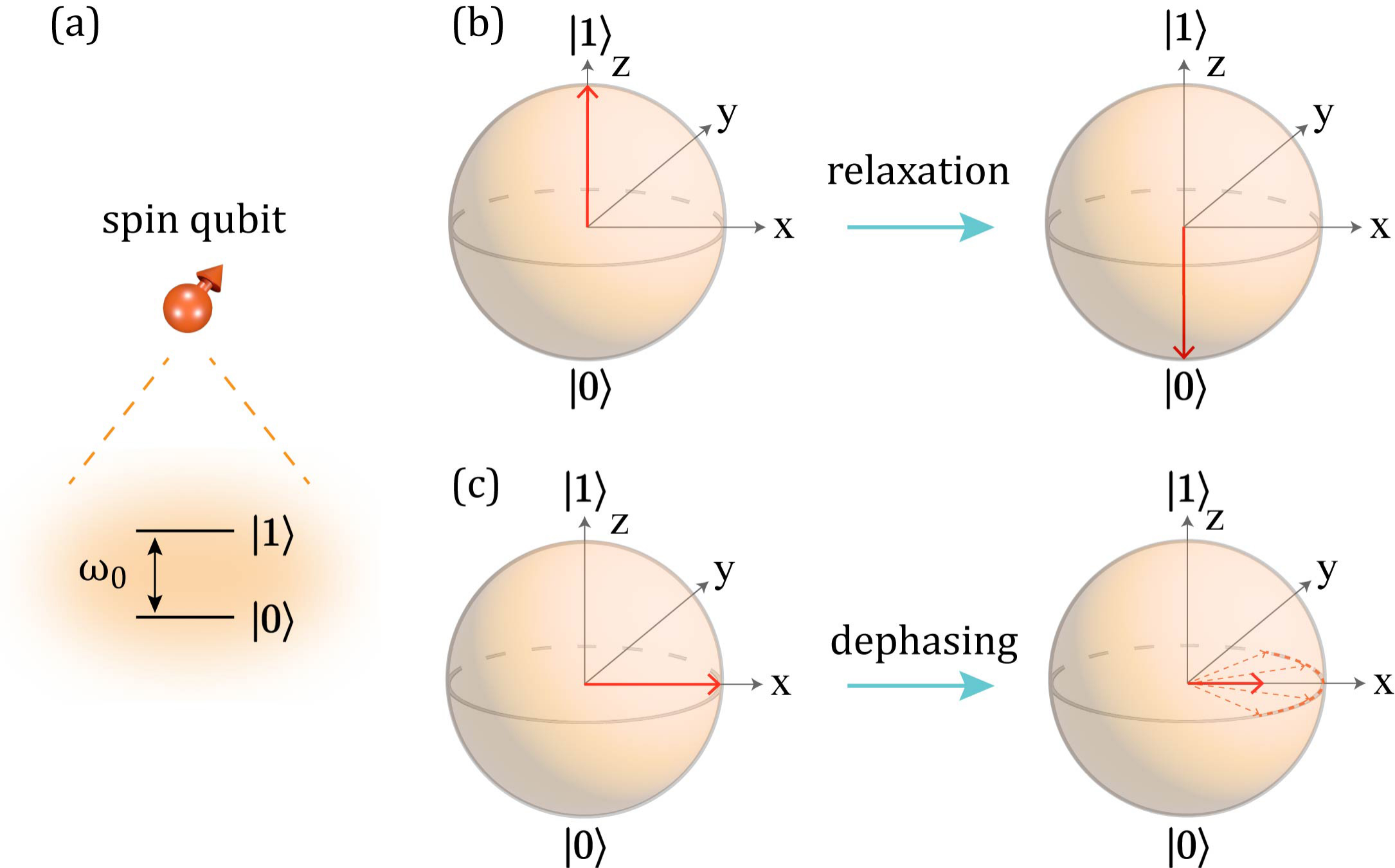}
    \caption{(a) A two-level system forms a quantum bit (qubit), which is the basic building block in quantum applications. The splitting frequency between the states is $\omega_0$. (b, c) Bloch sphere representation of (b) relaxation and (c) dephasing processes of the two-level system.}
    \label{fig:relaxation_dephasing}
\end{figure}

\subsection{Metrics for Quantifying Noise Effects}

We now discuss noise effects that lead to loss of quantum information and the metrics to quantify them. Here, we consider the spin qubit as a two-level system with splitting frequency $\omega_0$, as shown in Fig.~\ref{fig:relaxation_dephasing}(a).

\subsubsection{Relaxation}
In relaxation processes, quantum information encoded in qubits is lost due to energy exchange between qubits and the environment. In Fig.~\ref{fig:relaxation_dephasing}(b), we illustrate the relaxation processes in the Bloch sphere representation. Relaxation processes are usually induced by environmental noise at the qubit energy splitting frequency $\omega_0$ ($\omega_0\sim\mathrm{GHz}$ for spin qubits). The speed of relaxation processes can be characterized by the relaxation time $T_1$ or relaxation rates $\Gamma_r$.

\subsubsection{Pure dephasing}
A more prominent contribution to the loss of quantum information is the pure dephasing process. In pure dephasing, qubits lose their phase coherence without energy dissipation. In Fig.~\ref{fig:relaxation_dephasing}(c), we illustrate the pure dephasing processes in the Bloch sphere representation. Pure dephasing is usually induced by low-frequency noise ($\leq$MHz) not resonant with spin qubits and is usually faster than relaxation. The speed of pure dephasing processes can be characterized by the dephasing time $T_\phi$ or dephasing rates $\gamma_\phi$.

\subsubsection{Collective relaxation/dephasing}
In a multi-spin-qubit system, qubits can undergo collective relaxation and dephasing processes when the noise is correlated. The noise correlations are generally determined by the microscopic dynamics of the noise sources. In spin qubit quantum sensing and computing, noise induced by electromagnetic fluctuations can have strong correlations due to the prominent spatial correlations in near-field electromagnetic fluctuations~\cite{sun2024nano}. In quantum computing, collective relaxation/dephasing is detrimental to quantum error correction~\cite{klesse2005quantum}. Meanwhile, in quantum sensing, collective relaxation/dephasing can provide additional degrees of freedom to detect material properties. The speed of collective dephasing/relaxation processes can be characterized by the collective relaxation/dephasing rates.

\subsubsection{Gate fidelity}
In quantum computing, another important metric used to evaluate device performance is gate fidelity $F$. Noise in quantum computing systems can induce errors in quantum gate operations. Gate fidelity $F$ describes the closeness between physically implemented quantum gate operations and their theoretically ideal counterparts. Gate fidelity is usually closely related to $T_1$ and $T_\phi$, and is also determined by the gate operation time and gate operation protocol. Realizing higher gate fidelity can reduce the number of physical qubits needed to build a fault-tolerant quantum computer.

 \begin{figure*}
    \centering
    \includegraphics[width=6.4in]{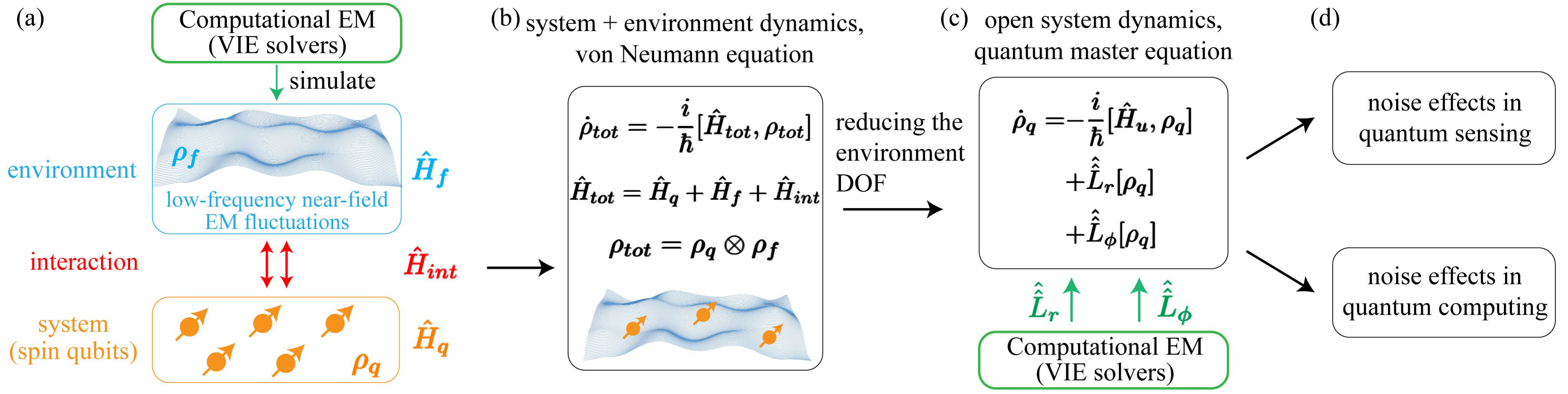}
    \caption{A schematic illustration of the quantum computational electromagentics framework for controlling noise effects in spin qubit quantum sensing and computing.}
    \label{fig:figureQCEM}
\end{figure*}

\section{Near-field Electromagnetic Fluctuations in Spin Qubit Quantum Applications}\label{section3}
In this section, we first discuss the origin of electromagnetic fluctuation noise in solid-state spin qubit systems. We highlight that in many quantum information applications, spin qubits are in a unique nano-electromagnetic environment, where low-frequency magnetic fluctuations are dominant and have strong spatial variations and correlations in the nanoscale range. Following this, we discuss the difficulties in modeling noise effects in the nano-electromagnetic environment.

Vacuum and thermal fluctuations of electromagnetic fields, which originate from the non-vanishing zero-point energy and the thermal photon bath, are universal phenomena that can occur at all frequencies and in all electromagnetic environments. Although their effects on macroscopic objects, such as Casimir forces and friction~\cite{buhmann2007dispersion}, can be very small, they can have a much more prominent influence on the properties of quantum objects. These universal electromagnetic fluctuations can induce noise effects in spin qubit systems, leading to loss of quantum information.

Electromagnetic fluctuations can be significantly modified in the near-field of materials. In many quantum information applications, spin qubits are in the vicinity of other components necessary for the control and reading out of quantum information. For example, in quantum computing devices based on semiconductor quantum-dot spin qubits, nanofabricated metallic/superconducting gates are widely used to control the interqubit interactions between spin qubits~\cite{takeda2021quantum,huang2019fidelity}. Metallic/superconducting antennas are employed to generate microwave pulse sequences to manipulate the spin qubit states. Nanomagnets are used to create magnetic field gradients to selectively control spin qubits~\cite{niknam2022quantum}. Similarly, quantum computing devices based on semiconductor donor spin qubits also employ metallic/superconducting gates and antennas to perform quantum gate operations~\cite{he2019two,morello2020donor,wang2016highly}. Meanwhile, in hybrid quantum systems, spintronic materials or superconducting resonators are coupled to spin qubits to realize long-range control~\cite{awschalom2021quantum} and quantum transduction~\cite{clerk2020hybrid}. Meanwhile, in quantum sensing applications, spin qubits are usually in the near-field of the material system to be probed~\cite{machado2023quantum,dolgirev2023local,kolkowitz2015probing,van2015nanometre}. 

One ubiquitous feature of the nano-electromagnetic environment near those nanostructured magnetic, metallic, and superconducting materials is the significant enhancement of low-frequency ($\leq \mathrm{GHz}$) magnetic field fluctuations~\cite{sun2024nano,khosravi2024giant}. At MHz and GHz frequencies, evanescent waves associated with intrinsic material loss can enhance the magnetic fluctuations $\langle \mathbf{B}(\mathbf{r},\omega) \mathbf{B}(\mathbf{r'},\omega)\rangle$ over 15 orders of magnitude compared to free space~\cite{sun2024nano}. These enhanced low-frequency near-field magnetic fluctuations can interact with spin qubits and induce relaxation (due to GHz magnetic field fluctuations) and dephasing (due to MHz magnetic field fluctuations) processes.

Another important feature of the low-frequency magnetic field fluctuations in the nano-electromagnetic environment is the prominent spatial dependence~\cite{sun2023limits}. In microwave cavities and free space, low-frequency electromagnetic fluctuations at two points separated by tens of nanometers do not change significantly. In contrast, low-frequency electromagnetic fluctuations can have stronger spatial dependence at different positions in the nano-electromagnetic environment due to evanescent wave contributions. In addition, low-frequency magnetic field fluctuations are usually spatially correlated in the nano-electromagnetic environment, leading to collective relaxation/dephasing processes~\cite{sun2024nano}. These jointly indicate the importance of accurately modeling low-frequency magnetic fluctuations in the nano-electromagnetic environment for controlling noise effects in spin qubit quantum sensing and computing.

\subsection{Difficulties in Modeling Noise in Nano-electromagnetic Environments}\label{section4a}
Despite its importance, controlling noise effects in the nano-electromagnetic environment is computationally challenging. Nanostructures in quantum devices can have a typical length scale $\sim 10\, \mathrm{nm}$. Meanwhile, the wavelength of electromagnetic fluctuations of interest can be larger than $1\, \mathrm{cm}$. Hence, modeling the nano-EM environment for spin qubits requires solving electrically small problems involving ultra-subwavelength objects. Solving these problems with differential equation based methods, such as FEM and FDTD methods, can lead to highly ill-conditioned numerical systems.

Furthermore, since low-frequency electromagnetic fluctuations in the nano-electromagnetic environment are dominated by evanescent wave contributions associated with intrinsic material loss, lossy materials can not be simply modeled by perfect electric/magnetic conductors~\cite{sun2024nano}. Instead, it is necessary to consider realistic material models that capture low-frequency microscopic dynamics in materials. Realistic materials can have large relative permittivity or permeability at low frequencies (e.g., metals). Therefore, the contrast ratio between the electromagnetic response of lossy materials and the environments can be very large. This increases the difficulty of modeling the nano-EM environment with perturbative methods, such as the Born-series-based iterations~\cite{van2020electromagnetic}, which can be difficult to converge for high contrast ratio~\cite{kleinman1990convergent}.

To overcome the above limitations, in this paper, we propose a quantum computational electromagnetics framework to control noise effects in spin qubit quantum computing and sensing. Our theoretical framework leverages advanced computational electromagnetics techniques to model fluctuations in the nano-electromagnetic environment. In the following, we first introduce the general formalism of our theoretical framework and then demonstrate its applications to spin qubit quantum computing and quantum sensing problems.

\section{Quantum Computational Electromagnetics Framework for Controlling Noise Effects in Spin Qubit Systems}\label{section4}

In this section, we present our numerical framework to control noise effects in spin qubit quantum applications. Based on foundations established in our previous works~\cite{sun2023limits,sun2024nano,cortes2022fundamental}, we first provide expressions for the electromagnetic fluctuation induced relaxation and dephasing effects in terms of the magnetic dyadic Green's function of the nano-electromagnetic environment. After this, we discuss how computational electromagnetics techniques, especially volume integral equation based methods, can efficiently calculate the dyadic Green's function necessary to model low-frequency magnetic fluctuations in quantum sensing and computing devices.

Here, we first describe the physical ideas underlying the theoretical framework. As shown in Fig.~\ref{fig:figureQCEM}(a), we start by considering a close system consisting of the near-field EM environment with density matrix $\rho_f$ and Hamiltonian $\hat{H}_f$, the spin qubit system with density matrix $\rho_q$ and Hamiltonian $\hat{H}_q$, and the interactions between the environment and the qubit system $\hat{H}_{int}$. Here, noise effects on spin qubits originate from the low-frequency near-field electromagnetic fluctuations, which can be accurately simulated by advanced computational electromagnetics, e.g., VIE based solvers. To study noise effects in spin qubit quantum applications, we need to understand the dynamics of spin qubits interacting with the electromagnetic environment. As shown in Fig.~\ref{fig:figureQCEM}(b), the dynamics of the total system (spin qubits + environment) can be described by the Liouville-von Neumann equation. However, directly solving the Liouville-von Neumann equation of the entire close system is generally complicated because the density matrix $\rho_f$ of the EM environment is infinite-dimensional. Considering that we are interested in the noise effects on spin qubit systems, it is not necessary to simulate the dynamics of the entire system. Instead, we can reduce the environment's degree of freedom (DOF) by tracing off the environment density matrix $\rho_f$ and only consider the dynamics of the open spin qubit systems. As shown in Fig.~\ref{fig:figureQCEM}(c), after this simplification, we obtain a quantum master equation~\cite{breuer2002theory} governing the dynamics of the spin qubit system. It is worth noting that the spin qubit system is an open system, and its dynamics are non-unitary. Noise effects from the low-frequency near-field EM environment are represented by the two non-unitary processes, relaxation $\hat{\hat{L}}_r$ and pure dephasing $\hat{\hat{L}}_\phi$. Computational electromagnetics can provide accurate simulations that determine $\hat{\hat{L}}_r$ and $\hat{\hat{L}}_\phi$ in the given nano-electromagnetic environment. As shown in Fig.~\ref{fig:figureQCEM}(d), by solving the quantum master equation for the spin qubit system, we can model and control noise effects in spin qubit quantum sensing and computing. 

In the following, we provide mathematical derivations following the above discussions. We consider an $N$-spin-qubit system coupled to electromagnetic fluctuations in the nano-electromagnetic environment. To facilitate the application of computational electromagnetics, we employ the quantization framework in macroscopic QED~\cite{buhmann2012macroscopic,scheel2009macroscopic}. We start from the total Hamiltonian for the $N$ spin qubits and the electromagnetic bath,
\begin{equation}\label{Htot}
    \hat{H}_{tot}=\hat{H}_q+\hat{H}_f+\hat{H}_{int},
\end{equation}
\begin{equation}
    \hat{H}_q=\sum_i^N \hbar \omega_{i} \hat{\sigma}_i^+ \hat{\sigma}_i^-,
\end{equation}
\begin{equation}
    \hat{H}_f=\int d^3 \mathbf{r} \int_0^\infty \hbar \omega \ \hat{\mathbf{f}}^\dagger (\mathbf{r},\omega) \hat{\mathbf{f}}(\mathbf{r},\omega),
\end{equation}
\begin{equation}\label{hint}
    \hat{H}_{int}=- \sum_i^N (\mathbf{m}_{i}^{eg}\hat{\sigma}_i^+ + \mathbf{m}_{i}^{ge}\hat{\sigma}_i^- + \mathbf{m}_{i}\hat{\sigma}_i^z) \cdot \hat{\mathbf{B}}(\mathbf{r_i}),
\end{equation}
where $\hat{H}_q$ and $\hat{H}_f$ are the Hamiltonians of the $N$-spin-qubit system and the electromagnetic bath, respectively. $\hat{H}_{int}$ describes the interactions between spin qubits and fluctuating magnetic fields. $\hat{\mathbf{B}}(\mathbf{r_i})$ is the magnetic field operator at position $\mathbf{r_i}$. $\hat{\mathbf{f}}^\dagger (\mathbf{r},\omega)$ and $\hat{\mathbf{f}}(\mathbf{r},\omega)$ are the field creation and annihilation operators. In Eq.~(\ref{hint}), we employ the point dipole approximation since the sizes of spin qubits are usually much smaller than the low-frequency electromagnetic field wavelength under consideration. $\omega_i$ and $\mathbf{r_i}$ are the splitting frequency and position of the i\textit{th} spin qubit. $\hat{\sigma}_i^+=|1\rangle \langle0|$, $\hat{\sigma}_i^-=|0\rangle \langle1|$, and $\hat{\sigma}_i^z=|1\rangle \langle1|-|0\rangle \langle0|$ represent the raising operator, lowering operator, and Pauli-$z$ operator for the i\textit{th} spin qubit. $\mathbf{m}_{i}^{eg}$, $\mathbf{m}_{i}^{ge}=\mathbf{m}_{i}^{eg\dagger}$, and $\mathbf{m}_{i}$ represent the spin magnetic moment of the i\textit{th} spin qubit perpendicular ($\mathbf{m}_{i}^{eg}$ and $\mathbf{m}_{i}^{ge}$) or parallel ($\mathbf{m}_{i}$) to the quantization axis. Here, noise effects arise due to the magnetic field fluctuations $\langle \hat{\mathbf{B}}(\mathbf{r_i}) \hat{\mathbf{B}}^\dagger(\mathbf{r_j})\rangle \neq 0$. Relaxation and pure dephasing processes are related to different terms in Eq.~(\ref{hint}). Relaxation processes are induced by interactions with transverse noise associated with $(\mathbf{m}_{i}^{eg}\hat{\sigma}_i^+ + \mathbf{m}_{i}^{ge}\hat{\sigma}_i^- ) \cdot \hat{\mathbf{B}}(\mathbf{r_i})$, while pure dephasing processes are induced by interactions with longitudinal noise associated with $ \mathbf{m}_{i}\hat{\sigma}_i^z \cdot \hat{\mathbf{B}}(\mathbf{r_i})$.

In macroscopic QED~\cite{buhmann2012macroscopic}, $\hat{\mathbf{B}}(\mathbf{r_i})$ is related to the dyadic Green's function $\overleftrightarrow{G}(\mathbf{r},\mathbf{r}',\omega)$ of the electromagnetic environment,
\begin{equation}
    \hat{\mathbf{B}}(\mathbf{r}) = \int_0^\infty d\omega [\hat{\mathbf{B}}(\mathbf{r},\omega)+\hat{\mathbf{B}}^\dagger(\mathbf{r},\omega)],
\end{equation}
\begin{equation}\label{Bfield}
    \hat{\mathbf{B}}(\mathbf{r},\omega) = (i \omega)^{-1} \int d^3 \mathbf{r}' \ \nabla_\mathbf{r} \times \overleftrightarrow{G}(\mathbf{r},\mathbf{r}',\omega) \cdot \hat{\mathbf{f}}(\mathbf{r}',\omega),
\end{equation}
and the field creation $\hat{\mathbf{f}}^\dagger (\mathbf{r},\omega)$ and annihilation $\hat{\mathbf{f}}(\mathbf{r},\omega)$ operators satisfy~\cite{yang2020single},
\begin{equation}
    [\hat{\mathrm{f}}_\alpha^\dagger(\mathbf{r},\omega), \hat{\mathrm{f}}_\beta(\mathbf{r}',\omega')]=\delta_{\alpha\beta}\delta(\mathbf{r}-\mathbf{r'})\delta(\omega-\omega'), 
\end{equation}
where $\hat{\mathrm{f}}_\alpha(\mathbf{r},\omega)$ is the component of $\hat{\mathbf{f}}(\mathbf{r},\omega)$ with $\alpha,\beta = x,y,z $. $\delta(\mathbf{r}-\mathbf{r'})$ and $\delta(\omega-\omega')$ are the Dirac delta functions. $\delta_{\alpha\beta}$ represents the Kronecker delta function.

From Eqs.~(\ref{hint}-\ref{Bfield}), we can see that interactions between spin qubits and the fluctuating electromagnetic fields are determined by the dyadic Green's function $\overleftrightarrow{G}(\mathbf{r},\mathbf{r}',\omega)$ of the electromagnetic environment, which is defined through,
\begin{equation}
    \nabla \times \nabla \times \overleftrightarrow{G}(\mathbf{r},\mathbf{r}',\omega) - k_0^2 \overleftrightarrow{G}(\mathbf{r},\mathbf{r}',\omega) = \overleftrightarrow{I} \delta (\mathbf{r}-\mathbf{r'}),
\end{equation}
where $k_0=\omega/c$ and $\overleftrightarrow{I}$ is the $3 \times 3$ identity matrix. 

Equations~(\ref{Htot}-\ref{hint}) govern the dynamics of the closed quantum system constituted of spin qubits and electromagnetic fields through the Liouville–von Neumann equation,
\begin{equation}\label{totliouville}
    \frac{d \rho_{tot}(t)}{d t}=\frac{1}{i \hbar} [\hat{H}_{tot},\rho_{tot}(t)],
\end{equation}
where the total density matrix $\rho_{tot}=\rho_{q} \otimes \rho_{f}$ is the Kronecker product of $N$-spin-qubit density matrix $\rho_q$ and electromagnetic field density matrix $\rho_f$. Directly solving Eq.~(\ref{totliouville}) can be complicated since $\rho_f$ is an infinite-dimensional matrix. Considering that we are interested in the noise effects on spin qubit systems, it is not necessary to simulate the closed quantum systems dynamics of the entire system. Instead, we can trace off the electromagnetic field part $\mathrm{Tr}_f [ \rho_{tot} ]=\rho_q$ on both sides of Eq.~(\ref{totliouville}) and only focus on the dynamics of the subsystem constituted of spin qubits. This subsystem is an open quantum system, which can have non-unitary time evolution in stark contrast to the closed quantum system. In this paper, we focus on the non-unitary component of the spin qubit subsystem dynamics, which represents the noise effects of electromagnetic fluctuations. 

For simplicity, we skip the algebraic steps involving tracing off the electromagnetic field components in Eq.~(\ref{totliouville}), which are similar to Refs.~\cite{sun2023limits,sun2024nano,cortes2022fundamental}. In the following, we provide our result for the open quantum systems dynamics of spin qubits. Here, we assume the weak-coupling condition, i.e., the electromagnetic bath is not significantly affected by spin qubit dynamics. This approximation is valid for the nano-electromagnetic environment, where photons from spin qubits can leave the local electromagnetic environment very fast. The $N$ spin qubits dynamics due to electromagnetic noise are given by,
\begin{equation}\label{me}
    \frac{d \rho_q(t)}{d t} = -\frac{i}{\hbar}[\hat{H}_{u}(t),\rho_q(t)] + \hat{\hat{L}}_r[\rho_q(t)] + \hat{\hat{L}}_\phi[\rho_q(t)],
\end{equation}
where $\hat{H}_{u}$ is the unitary evolution Hamiltonian, which includes the dipole-dipole interaction Hamiltonian mediated by the electromagnetic bath, Hamiltonian for other types of interactions between spin qubits (e.g., exchange interactions), and Hamiltonian for control pulse sequence. $\hat{\hat{L}}_r$ and $\hat{\hat{L}}_\phi$ represent superoperators describing the relaxation and pure dephasing processes of the spin qubit system. The first term on the right-hand side of Eq.~(\ref{me}) represents the unitary evolution of the $N$-spin-qubit system. The last two terms represent the non-unitary evolution induced by the electromagnetic noise. 

In the following, we focus on the $\hat{\hat{L}}_r[\rho_q(t)] $ and $\hat{\hat{L}}_\phi[\rho_q(t)]$ terms. We provide the expressions of $\hat{\hat{L}}_r$ and $\hat{\hat{L}}_\phi$ in terms of the magnetic dyadic Green's function $\overleftrightarrow{G}_m(\mathbf{r},\mathbf{r}',\omega)$, which can be efficiently modeled by volume integral equation methods, as discussed in section.~\ref{section_cem}. The magnetic dyadic Green's function $\overleftrightarrow{G}_m(\mathbf{r},\mathbf{r}',\omega)$ is defined as,
\begin{equation}
    \overleftrightarrow{G}_m (\mathbf{r,r'},\omega) = \frac{1}{k_0 ^2} \nabla \times \overleftrightarrow{G}(\mathbf{r,r'},\omega) \times \nabla',
\end{equation}
where the components of $\overleftrightarrow{G}_m (\mathbf{r,r'},\omega)$ can be expressed with the Levi-Civita symbols $\epsilon_{\alpha kl}$ and $\epsilon_{\beta nm}$ as~\cite{sun2023limits}:
\begin{equation}
    \left[\overleftrightarrow{G}_m (\mathbf{r,r'},\omega)\right]_{\alpha \beta}=\epsilon_{\alpha kl}\epsilon_{\beta nm}\partial^k\partial^{'n} \left[\overleftrightarrow{G}_m (\mathbf{r,r'},\omega)\right]^{lm}.
\end{equation}

\subsection{Modeling Spin Qubit Relaxation}
The relaxation processes in spin qubit systems are induced by transverse noise associated with $\mathbf{m}_{i}^{eg}\hat{\sigma}_i^+ + \mathbf{m}_{i}^{ge}\hat{\sigma}_i^-$ terms in Eq.~(\ref{hint}) and are related to GHz noise resonant with qubit frequencies $\omega_i$. $\hat{\hat{L}}_r[\rho_q]$ in Eq.~(\ref{me}) is~\cite{sun2023limits}
\begin{align}
\begin{aligned}\label{Lr}
    &\hat{\hat{L}}_r[\rho_q]    =\\
    &(\bar{\mathcal{N}}+1) \sum_{i,j}^N  \Gamma_r^{ij}  \Big[\hat{\sigma}_i^- \rho_q(t) \hat{\sigma}_j^+ 
    - \frac{1}{2} \rho_q \hat{\sigma}_i^+ \hat{\sigma}_j^- - \frac{1}{2} \hat{\sigma}_i^+ \hat{\sigma}_j^- \rho_q\Big] \\
    &+ \bar{\mathcal{N}} \sum_{i,j}^N \Gamma_r^{ij}  \Big[\hat{\sigma}_i^+ \rho_q \hat{\sigma}_j^- 
    - \frac{1}{2} \rho_q \hat{\sigma}_i^- \hat{\sigma}_j^+ - \frac{1}{2} \hat{\sigma}_i^- \hat{\sigma}_j^+ \rho_q\Big],
\end{aligned}
\end{align}
where $\bar{\mathcal{N}}$ is the mean photon number determined by the temperature $\mathcal{T}_{em}$ of the electromagnetic bath,
\begin{equation}
    \bar{\mathcal{N}}=\frac{1}{e^{\hbar \omega/k_b \mathcal{T}_{em}}-1},
\end{equation}
and $\Gamma_r^{ij}$ represent the relaxation rates, which determine the energy dissipation of the quantum system.

The electromagnetic noise induced single-qubit relaxation rate ($i=j$) is proportional to the one-point dyadic Green's function~\cite{sun2023limits},
\begin{equation}\label{rel_rate}
    \Gamma_r^{ii} = \frac{2\mu_0 k_0^2}{\hbar}   \mathbf{m}_{i}^{eg}  \cdot  \mathrm{Im} \, \overleftrightarrow{G}_m  (\mathbf{r}_i,\mathbf{r}_i,\omega_i)  \cdot \mathbf{m}_{i}^{ge}.
\end{equation}

Meanwhile, electromagnetic noise is correlated. This leads to the correlated relaxation between two spin qubits at rates $\Gamma_r^{ij}$ proportional to the two-point dyadic Green's function~\cite{sun2023limits},
\begin{equation}\label{crel_rate}
    \Gamma_r^{ij} = \frac{2\mu_0 k_0^2}{\hbar}   \mathbf{m}_{i}^{eg} \cdot \mathrm{Im} \, \overleftrightarrow{G}_m  (\mathbf{r}_i,\mathbf{r}_j,\omega_+)   \cdot \mathbf{m}_{j}^{ge},
\end{equation}
where $\omega_+=(\omega_i+\omega_j)/2$ and we have assumed $|\omega_i-\omega_j|\ll\omega_i+\omega_j$. 

For spin qubits, the energy splitting frequency is usually around the GHz range. Therefore, Eqs.~(\ref{rel_rate}) and (\ref{crel_rate}) connect the relaxation rates of the spin qubit systems to the GHz magnetic dyadic Green's functions, which can be accurately calculated by the volume integral equation methods discussed in the next section. 

\subsection{Modeling Spin Qubit Dephasing}
The pure dephasing processes in spin qubit systems are induced by longitudinal electromagnetic noise associated with the $\mathbf{m}_{i}\hat{\sigma}_i^z$ term in Eq.~(\ref{hint}) and are related to low-frequency ($\leq$MHz) noise not resonant with spin qubits. $\hat{\hat{L}}_\phi[\rho_q]$ in Eq.~(\ref{me}) is~\cite{sun2024nano}
\begin{equation}\label{Lphi}
    \hat{\hat{L}}_\phi[\rho_q]  = \sum_{i,j}^N   \gamma_\phi^{ij}(t) \Big[\hat{\sigma}_{i}^z \rho_q \, \hat{\sigma}_{j}^z-\frac{1}{2} \hat{\sigma}_{i}^z \hat{\sigma}_{j}^z \rho_q - \frac{1}{2} \rho_q \, \hat{\sigma}_{i}^z \hat{\sigma}_{j}^z \Big],
\end{equation}
where $\gamma_\phi^{ij}$ represent the pure dephasing rates.

The electromagnetic noise induced single-qubit dephasing rate ($i = j$) is determined by the frequency integral of low-frequency one-point dyadic Green’s function~\cite{sun2024nano},
\begin{multline}\label{dep_rate}
    \gamma_\phi^{ii}(t) = \frac{4\mu_0}{\hbar \pi} \int_0^{\omega_c} d\omega \ \frac{\sin{\omega t}}{\omega} (\bar{\mathcal{N}}+\frac{1}{2}) \, \frac{\omega^2}{c^2} \\ \mathbf{m}_{i} \cdot  \mathrm{Im}\overleftrightarrow{G}_m(\mathbf{r_i},\mathbf{r_i},\omega)  \cdot \mathbf{m}_{i},
\end{multline}
where $\omega_c$ is the cutoff frequency necessary for the convergence of Eq.~(\ref{dep_rate})~\cite{sun2024nano}. 

Meanwhile, correlations in electromagnetic noise lead to the collective dephasing between two
spin qubits at rates $\gamma_\phi^{ij}(t)$ proportional to the frequency integral of two-point dyadic Green’s function~\cite{sun2024nano},
\begin{multline}\label{cdep_rate}
    \gamma_\phi^{ij}(t) = \frac{4\mu_0}{\hbar \pi} \int_0^{\omega_c} d\omega \ \frac{\sin{\omega t}}{\omega} (\bar{\mathcal{N}}+\frac{1}{2}) \, \frac{\omega^2}{c^2} \\ \mathbf{m}_{i} \cdot  \mathrm{Im}\overleftrightarrow{G}_m(\mathbf{r_i},\mathbf{r_j},\omega) \cdot \mathbf{m}_{j}.
\end{multline}

The integral in Eqs.~(\ref{dep_rate}) and (\ref{cdep_rate}) is usually dominated by contributions from MHz frequencies. Therefore, Eqs.~(\ref{dep_rate}) and (\ref{cdep_rate}) connect the dephasing rates of the spin qubit systems to the MHz magnetic dyadic Green's functions, which can also be accurately modeled by the volume integral equation methods discussed in the next section.

\subsection{Modeling Quantum Gate Infidelity}
From the above equations, we can simulate the open quantum systems dynamics with the $\mathrm{Im} \, \overleftrightarrow{G}_m$ calculated from computational electromagnetics. We now discuss how to model the infidelity of quantum gate operations induced by electromagnetic noise. For given initial density matrix $\rho_q(0)$, the final density matrix $\rho_q(\rho_q(0),t_f)$ when the quantum gate operation is completed can be calculated from Eqs.~(\ref{rel_rate})-(\ref{cdep_rate})~\cite{sun2023limits}. The fidelity $F$ of a given quantum gate operation can be defined as,
\begin{equation}\label{fidelity}
    F=\sum_{\rho_q(0)} \mathrm{Tr} \, [\rho_q(\rho_q(0),t_f) \, \tilde{\rho}_q(\rho_q(0),t_f)]/\sum_{\rho_q(0)} 1,
\end{equation}
where $\tilde{\rho}_q(\rho_q(0),t_f)$ is the ideal final density matrix when a perfect non-noisy quantum gate is applied~\cite{sun2023limits}. Equation~(\ref{fidelity}) involves an average over different pure initial states $\rho_q(0)$, which can be selected as the four computational basis states $|00\rangle, |01\rangle, |10\rangle, |11\rangle$ for two-qubit gates. Meanwhile, $F$ averaged over over all input states can be calculated from the average over orthogonal basis of $\mathcal{D} \times \mathcal{D}$ unitary operators~\cite{nielsen2002simple}, where $\mathcal{D}$ is the dimension of the quantum system (e.g., $\mathcal{D}=2^N$ for a $N$ qubit system). The quantum gate infidelity $\Delta F$ can be calculated from the difference in $F$ when the electromagnetic fluctuation noise (i.e., Eqs.~(\ref{Lr}) and (\ref{Lphi})) is included or neglected in simulating the quantum dynamics Eq.~(\ref{me}).

To this end, we briefly discuss how to apply computational electromagnetics in modeling noise effects for quantum sensing and computing. In quantum sensing, material properties are probed by the change of the relaxation $\Gamma_r$ and dephasing rates $\gamma_\phi$ when the spin qubits are in the vicinity or away from the materials. In quantum computing, the relaxation rates $\Gamma_r$, dephasing rates $\gamma_\phi$, and quantum gate fidelity characterize the device performance. From the above discussions, we can find that these quantities are determined by the magnetic dyadic Green's function of the nano-electromagnetic environment, which can be efficiently modeled by volume integral equation based methods discussed in the next section.

\section{Volume Integral Equation Methods}\label{section_cem}
In this section, we first briefly discuss the current approximate method used for modeling $\overleftrightarrow{G}_m$ and near-field magnetic fluctuations in spin qubit quantum sensing and computing applications. We demonstrate that this method is not valid in the near-field of nanostructures or inhomogeneous materials. Then, we show how volume integral equation methods can be used to accurately calculate the magnetic dyadic Green's function of the nano-electromagnetic environment. 

In the near-field of material interfaces, the low-frequency magnetic dyadic Green's function is dominated by contributions from the reflected component,
\begin{align}
\begin{aligned}
    \overleftrightarrow{G}_m (\mathbf{r}_i,\mathbf{r}_j,\omega) = &\overleftrightarrow{G}_m^0 (\mathbf{r}_i,\mathbf{r}_j,\omega)+\overleftrightarrow{G}_m^r (\mathbf{r}_i,\mathbf{r}_j,\omega) \\
    \approx &\overleftrightarrow{G}_m^r (\mathbf{r}_i,\mathbf{r}_j,\omega),
\end{aligned}  
\end{align}
since the free space component is negligible at low frequencies $\overleftrightarrow{G}^0_m (\mathbf{r}_i,\mathbf{r}_j,\omega) \approx \omega/6\pi c$.

One commonly used approximation to model low-frequency magnetic fluctuations in quantum sensing and computing devices is to assume that the materials have a planar geometry~\cite{sun2023limits,poudel2013relaxation}. In this case, $ \overleftrightarrow{G}_m^r (\mathbf{r}_i,\mathbf{r}_j,\omega)$ can be calculated from the four Fresnel reflection coefficients $r_{ss}$, $r_{sp}$, $r_{ps}$, and $r_{pp}$ through~\cite{khandekar2019thermal},
\begin{align}
\begin{aligned}\label{anacpd}
    \overleftrightarrow{G}^r_m (\mathbf{r}_i,&\mathbf{r}_j,\omega)=\frac{i}{8 \pi^2}\int \frac{d \mathbf{q}}{k_z} e^{i \mathbf{q}(\mathbf{r}_i-\mathbf{r}_j)} e^{i k_z(z_i+z_j)} \\ 
    & \biggl( \frac{r_{pp}}{q^2} \begin{bmatrix} q_y^2&-q_x q_y&0\\-q_x q_y&q_x^2&0\\0&0&0 \end{bmatrix}\\ 
    &+ \frac{r_{ss}}{k_0^2q^2}\begin{bmatrix} -q_x^2 k_z^2 & -q_x q_y k_z^2 & -q_x k_z q^2\\-q_x q_y k_z^2 & -q_y^2 k_z^2 & -q_y k_z q^2\\q^2 q_x k_z & q^2 q_y k_z & q^4 \end{bmatrix} \\
    & + \frac{r_{ps}}{k_0 q^2}\begin{bmatrix} q_x q_y k_z & q_y^2 k_z & q_y q^2\\-q_x^2 k_z & -q_y q_x k_z & -q_x q^2\\0 & 0 & 0 \end{bmatrix} \\ 
    &+ \frac{r_{sp}}{k_0 q^2}\begin{bmatrix} -q_x q_y k_z & q_x^2 k_z & 0\\-q_y^2 k_z & q_y q_x k_z & 0\\ q_y q^2 & - q_x q^2 & 0 \end{bmatrix} \biggr),
\end{aligned}
\end{align}
where $\mathbf{q}=q_x \hat{\mathbf{x}} + q_y \hat{\mathbf{y}}$ is the in-plane momentum, $q=|\mathbf{q}|$ is the magnitude of $\mathbf{q}$, $k_z=\sqrt{k_0^2-q^2}$ is the $z$-component of the momentum, $z_i$ and $z_j$ are the $z$ components of $\mathbf{r_i}$ and $\mathbf{r_j}$.

However, in the near-field nanostructures or inhomogeneous materials, the reflection coefficients become ill-defined, and Eq.~(\ref{anacpd}) is not valid for calculating $ \overleftrightarrow{G}_m^r (\mathbf{r}_i,\mathbf{r}_j,\omega)$. Meanwhile, nanostructures and inhomogeneous materials are commonly encountered in quantum sensing, e.g., in nanoscale imaging of material properties. Additionally, in quantum computing devices, metallic/superconducting gates are usually nanostructured to control the properties of spin qubits. Therefore, in the next subsection, we discuss the volume integral equations based method to model noise effects near arbitrarily structured materials.

\subsection{Volume Integral Equations}
We first notice that the magnetic dyadic Green's function connects a magnetic dipole $\mathbf{m}(\mathbf{r}',\omega)$ to the magnetic field $\mathbf{H}(\mathbf{r},\omega)$ through,
\begin{align}
\begin{aligned}\label{magneticG}
    \mathbf{H}(\mathbf{r},\omega)=k_0^2 \, \overleftrightarrow{G}_m (\mathbf{r,r'},\omega) \cdot \mathbf{m}(\mathbf{r}',\omega)
\end{aligned}
\end{align}
Therefore, $\overleftrightarrow{G}_m^r $ can be obtained from the scattered fields of a magnetic dipole in the vicinity of nanostructured materials.

In the following, we solve the scattered fields of a magnetic dipole by using volume integral equations. We first consider an arbitrarily structured medium with complex permittivity $\varepsilon_m$ occupying a region $V$ in the space. Inside the medium, we have $\mathbf{D}(\mathbf{r})=\varepsilon_m(\mathbf{E}^i(\mathbf{r}) + \mathbf{E}^{s}(\mathbf{r}))$, where $\mathbf{E}^i(\mathbf{r})$ is the incident electric field from the magnetic dipole and $\mathbf{E}^{s}(\mathbf{r})$ is the scattered electric field from the nanostructures. Under the Lorentz gauge and assuming the $e^{j\omega t}$ time dependence, we have,
\begin{align}
\begin{aligned}\label{vie1}
    \mathbf{E}^{s}(\mathbf{r})=& -j\omega \mathbf{A}(\mathbf{r})-\nabla \phi(\mathbf{r}) \\
    =& -j\omega \mu_0 \int_V\mathbf{J}_b(\mathbf{r'})g(\mathbf{r},\mathbf{r'})d\mathbf{r'} \\
    &-\frac{1}{\varepsilon_0} \int_V\rho_b(\mathbf{r'})\nabla g(\mathbf{r},\mathbf{r'})d\mathbf{r'}
\end{aligned}
\end{align}
where $\mathbf{J}_b$ and $\rho_b$ are the bound currents and charge densities. $g(\mathbf{r},\mathbf{r'})=e^{-jk_0\|\mathbf{r}-\mathbf{r'}\|}/4\pi \|\mathbf{r}-\mathbf{r'}\|$ is the scalar Green's function. From the volume equivalent principle and constitutive relations, we have $\rho_b(\mathbf{r})=-\nabla \cdot (\kappa(\mathbf{r})\mathbf{D}(\mathbf{r}))$ and $\mathbf{J}_b(\mathbf{r})=j \omega  (\kappa(\mathbf{r})\mathbf{D}(\mathbf{r}))$. $\kappa(\mathbf{r})=(\varepsilon_m-\varepsilon_0)/\varepsilon_m$ is the contrast ratio between the permittivity of the media and the surrounding space. Substituting these two equations into Eq.~(\ref{vie1}), we obtain the following volume integral equation in terms of the electric flux density $\mathbf{D}(\mathbf{r})$,
\begin{multline}\label{vie2}
    \mathbf{E}^i(\mathbf{r}) = \mathbf{E}(\mathbf{r}) - \mathbf{E}^{s}(\mathbf{r}) = \frac{\mathbf{D}(\mathbf{r})}{\varepsilon(\mathbf{r})} - \int_V \Big \{ \mu_0 \omega^2 \kappa (\mathbf{r'}) \\ \mathbf{D}(\mathbf{r'}) + \frac{1}{\varepsilon_0} \nabla' \cdot ( \kappa (\mathbf{r'}) \mathbf{D}(\mathbf{r'})  ) \nabla \Big \} \ g(\mathbf{r}, \mathbf{r'})\ d \mathbf{r}'.
\end{multline}

To solve the volume integral equation~(\ref{vie2}), we first discretize the volume $V$ into a tetrahedral mesh to model arbitrarily shaped geometry. Then, we expand the electric flux density $\mathbf{D}(\mathbf{r})$ using the Schaubert-Wilton-Glisson (SWG) basis functions in each tetrahedral element and apply a standard Galerkin testing. This converts Eq.~(\ref{vie2}) into a matrix equation. For a small number of unknowns, the matrix equation can be solved directly. Meanwhile, for a large number of unknowns, we can employ fast solvers to represent the dense matrix by a reduced set of parameters and solve it efficiently~\cite{2018_Maiomiao_Direct,omar2015linear,YifanWang_TAP2022}. 

After solving Eq.~(\ref{vie2}), we can obtain the scattered electric fields $\mathbf{E}^{s}(\mathbf{r})$ and bound currents $\mathbf{J}_b(\mathbf{r})$. We can then find the scattered magnetic fields $\mathbf{H}^s(\mathbf{r})$ through,
\begin{equation}\label{Hr}
    \mathbf{H}^s(\mathbf{r}) = -\frac{1}{j\omega \mu_0} \nabla \times \mathbf{E}^s(\mathbf{r})
    = \int_V \nabla g(\mathbf{r}, \mathbf{r'}) \times \mathbf{J}_b(\mathbf{r}) d \mathbf{r}'.
\end{equation}
With $\mathbf{H}^s(\mathbf{r})$ and Eq.~(\ref{magneticG}), we can obtain the $3 \times 3$ magnetic dyadic Green's function $\overleftrightarrow{G}_m^r$ necessary to model noise effects in the nano-electromagnetic environment in quantum sensing and computing.

To this end, we have established a numerical framework to model low-frequency magnetic noise in the nano-electromagnetic environment. In the following two sections, we demonstrate how to apply our theoretical framework in controlling noise effects in realistic quantum computing and sensing applications.

\section{Controlling Noise Effects in Spin Qubit Quantum Computing}\label{section_qc}

In this section, we apply our theoretical framework to model noise effects in spin qubit quantum computing devices. We employ volume integral equation methods (section~\ref{section_cem}) to calculate single-qubit errors and correlated errors induced by near-field magnetic fluctuations $\mathrm{Im} \, \overleftrightarrow{G}_m(\mathbf{r}_i,\mathbf{r}_i,\omega)$ and fluctuation correlations $\mathrm{Im} \, \overleftrightarrow{G}_m(\mathbf{r}_i,\mathbf{r}_j,\omega)$. We demonstrate that in realistic devices, noise and noise correlations can exhibit behaviors qualitatively different from the approximate results predicted by Eq.~(\ref{anacpd}).

Here, we study the low-frequency magnetic field fluctuations in a state-of-the-art quantum computing device reported in Ref.~\cite{takeda2021quantum}, which contains three semiconductor spin qubits based on silicon/silicon–germanium heterostructure. In this device, top aluminum gates are employed for the initialization, control, and readout of the three spin qubits. In Fig.~\ref{fig:figure1}(a), we demonstrate a schematic of the simplified device structure considered in our volume integral equation simulations. In our calculations, we consider the spin qubit frequency to be $18$GHz~\cite{takeda2021quantum} and the aluminum conductivity $\sigma_{Al}=1.6\times 10^8\,\mathrm{S/m}$~\cite{de1988temperature} at low temperatures ($< 1\,\mathrm{K}$) corresponding to the operation temperature of the quantum computing device. We consider the total length and width of the metallic gates to be $1200\,\mathrm{nm}$ and $650\,\mathrm{nm}$, and the thickness of the metallic gates to be $100\,\mathrm{nm}$ for the thinner regions and $150\,\mathrm{nm}$ for the thicker regions, as shown in Fig.~\ref{fig:figure1}(a).

To exemplify the application of volume integral equation methods in controlling noise and noise correlation effects, we study the single-qubit $\Gamma_r^{ii}$ (Eq.~(\ref{rel_rate})) and correlated relaxation rates $\Gamma_r^{ij} (i \neq j)$ (Eq.~(\ref{crel_rate})) of spin qubits near the aluminum gates. Relaxation processes involve quantum state transitions and are the dominant source of leakage error~\cite{wood2018quantification}. Other noise effects, including pure dephasing and quantum gate infidelity, can be modeled similarly by volume integral equation methods. In addition, we compare the relaxation rates modeled by volume integral equation methods and their counterparts calculated from the approximate thin film solutions (Eq.~(\ref{anacpd})) to demonstrate the limitation of approximate solutions for noise mitigation in real devices.

In Fig.~\ref{fig:figure1}(b), we demonstrate the tetrahedral mesh used to discretize the metal gates and solve the VIEs. Here, we employ 53525 tetrahedron elements with refinement in the region close to the spin qubit positions to improve the accuracy of VIE simulations. The corresponding VIE system matrix in the calculations takes around 197 GB of memory. The solution time using a 128-core workstation is around 1156 seconds for one spatial configuration of spin qubits.

\subsection{Single-qubit Error}

\begin{figure}[!t]
    \centering
    \includegraphics[width=3in]{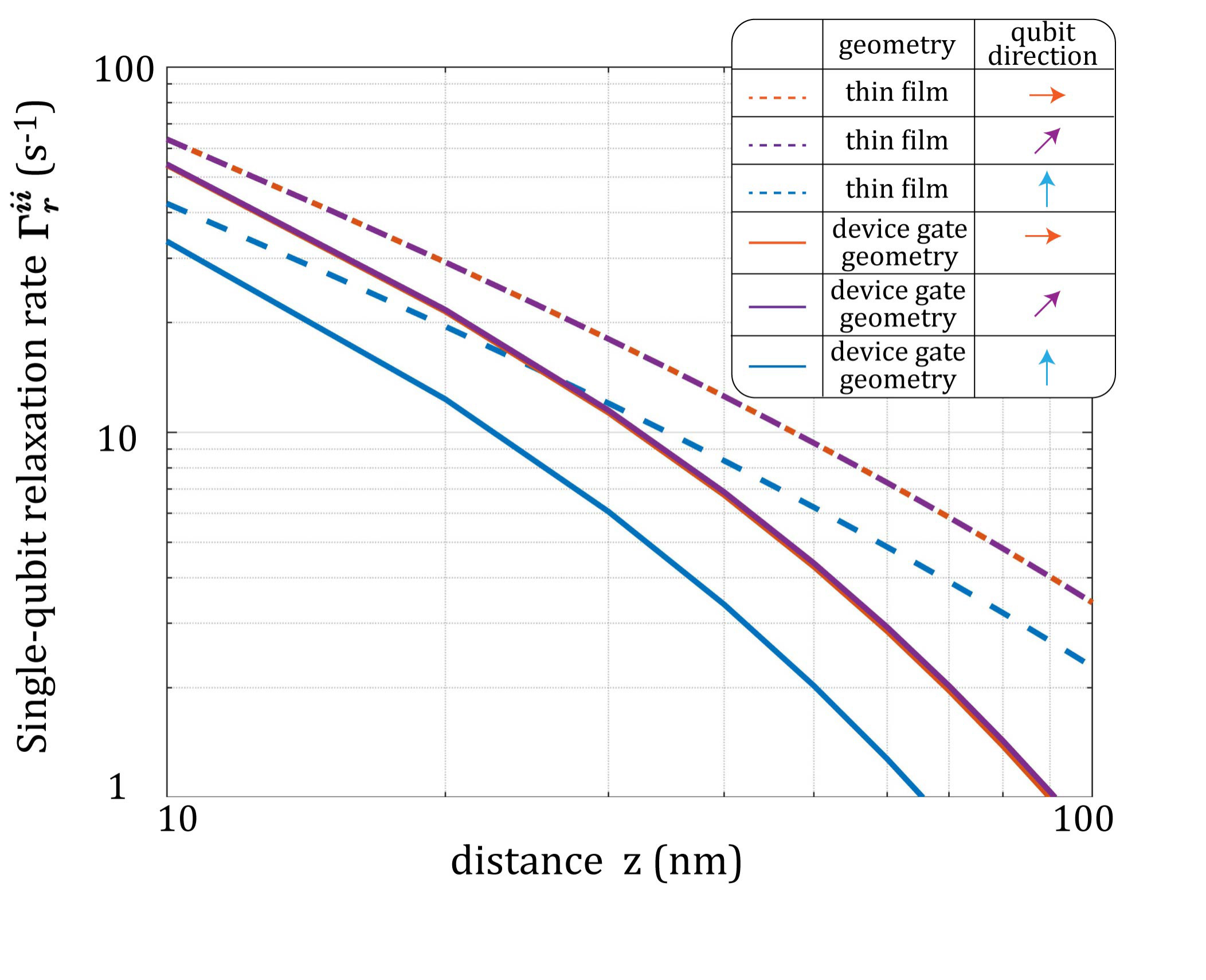}
    \caption{Modeling single-qubit error with volume integral equation based methods. We demonstrate the single-qubit relaxation rate $\Gamma^{ii}_r$ of a spin qubit at a distance $z$ from aluminum thin films (dashed curves) and aluminum gates with device gate geometry (solid curves) shown in Fig.~\ref{fig:figure1}. We compare $\Gamma_r^{ii}$ for spin qubits with quantization axis along the x direction (orange curves), y direction (purple curves), and z direction (blue curves). The thin film approximation is not valid for modeling electromagnetic noise at large $z$.}
    \label{fig:figure3}
\end{figure}

In Fig.~\ref{fig:figure3}, we present the single-qubit relaxation rates $\Gamma^{ii}_r$ for the center spin qubit at distance $z$ ranging from $10$ to $100\,\mathrm{nm}$ away from metallic gates (see Fig.~\ref{fig:figure1}). We compare $\Gamma^{ii}_r$ modeled by volume integral equation methods near metal gates with realistic device gate geometry (solid curves) and $\Gamma^{ii}_r$ calculated from approximate thin film solutions near an aluminum thin film of $125\,\mathrm{nm}$ thickness (dashed curves). We consider three scenarios where the spin qubit quantization axis is along the x direction (orange curves), y direction (purple curves), and z direction (blue curves). Here, the thin film approximation is valid only when the spin qubits are sufficiently close to the metallic contacts. This matches with our observations in Fig.~\ref{fig:figure3}, where we can find that the thin film approximation results $\Gamma^{ii}_r$ deviate significantly from realistic noise effects at large $z$ and start to approach $\Gamma^{ii}_r$ modeled by volume integral equation methods only at small $z<10\,\mathrm{nm}$. However, significant quantitative differences still exist even at small $z$. Additionally, we can find that $\Gamma^{ii}_r$ modeled by volume integral equation methods exhibits a scaling law concerning $z$ qualitatively different from the approximate results, which follow the simple $\Gamma^{ii}_r \sim 1/z$. Furthermore, we can also find that the ratio between relaxation rates $\Gamma^{ii}_r$ of spin qubits with different quantization axes is different near device gate geometry and thin films. The above results indicate that the approximate results based on Eq.~(\ref{anacpd}) may not predict qualitatively correct noise behaviors in realistic devices, and it is important to employ the volume integral equation based methods for noise analysis and device engineering to improve the performance of spin qubit quantum computing devices.

\subsection{Correlated Error}

\begin{figure}[!t]
    \centering
    \includegraphics[width=3in]{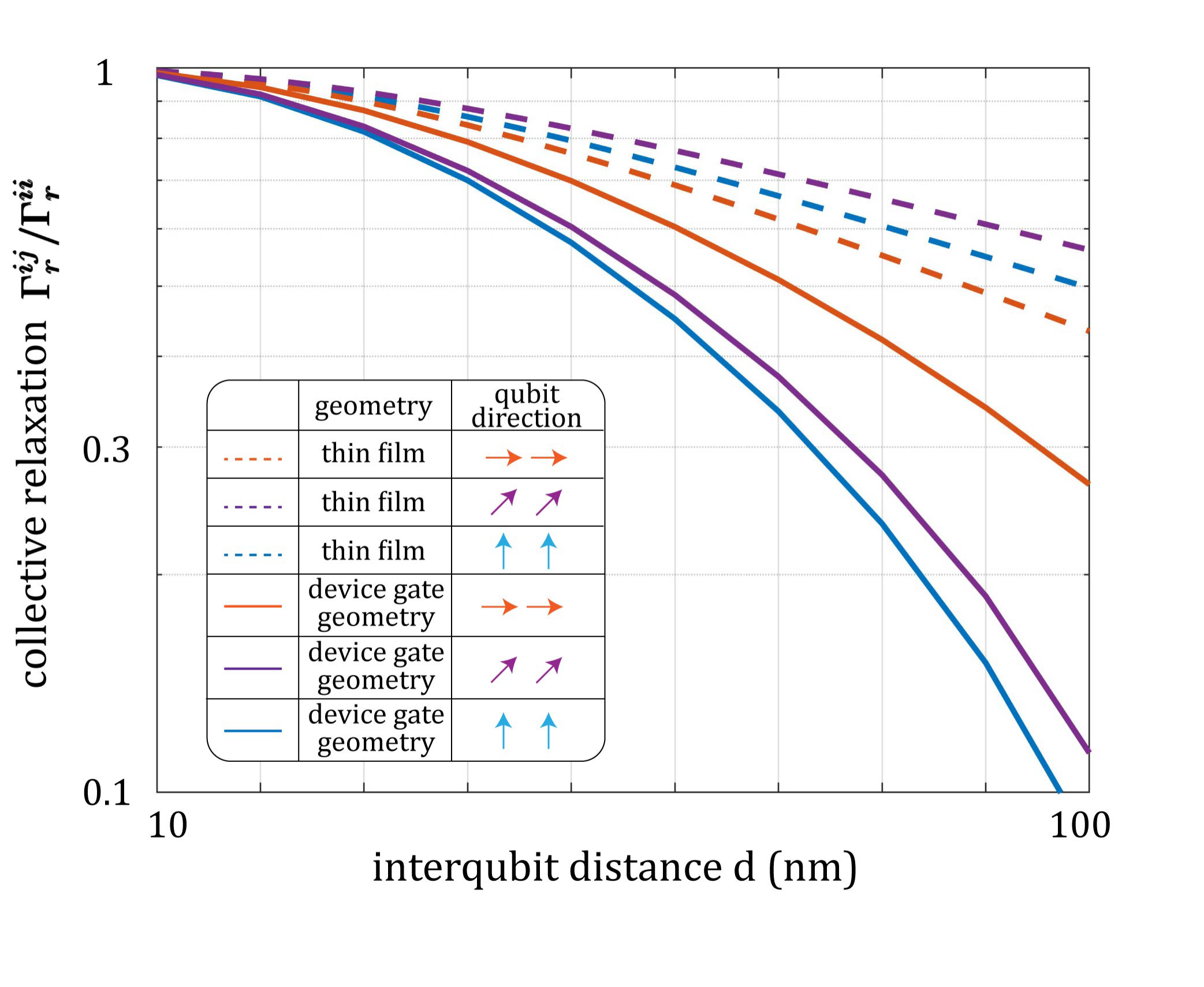}
    \caption{Modeling correlated error with volume integral equation based methods. We demonstrate the ratio of correlated and single-qubit relaxation rates $\Gamma_r^{ij}/\Gamma_r^{ii}$ for two spin qubits with interqubit distance $d$ at 40 nm from aluminum thin films (dashed curves) and aluminum gates with device gate geometry shown in Fig.~\ref{fig:figure1} (solid curves). We compare $\Gamma_r^{ij}/\Gamma_r^{ii}$ for spin qubits with quantization axis along the x direction (orange curves), y direction (purple curves), and z direction (blue curves). The thin film approximation is not valid for modeling electromagnetic noise correlations at large $d$.}
    \label{fig:figure4}
\end{figure}

In Fig.~\ref{fig:figure4}, we demonstrate the correlated relaxation rates $\Gamma^{ij}_r$ for two spin qubits at distance $z=40\,\mathrm{nm}$ from an aluminum thin film (dashed curves) and aluminum gates with device gate geometry (solid curves). We present the ratio between correlated and single-qubit relaxation rates $\Gamma_r^{ij}/\Gamma_r^{ii}$, which indicates the correlation range of low-frequency magnetic fluctuations in devices. We consider one qubit is at the center qubit position, and the other is separated by interqubit distance $d$ ranging from $10$ to $100\,\mathrm{nm}$ along the x direction (see Fig.~\ref{fig:figure1}). We compare the results when the quantization axes of both spin qubits are along the x direction (orange curves), y direction (purple curves), and z direction (blue curves). We can find that the thin film approximation is valid only when the two qubits are very close to each other with small interqubit distance $d$ and deviates significantly from the volume integral equation based results at large $d$. This can be understood by considering that when two spin qubits are separated by larger d, metallic contacts perceived by the spin qubits are not uniform thin films, and the nano-structures of metallic contacts play an important role in determining the EM noise correlations. Furthermore, we can find that near realistic metal gates, the noise correlation range is generally suppressed compared to the thin film results. 
Meanwhile, the suppression is sensitive to the directions of spin qubits. We can see that the correlated relaxation is of a longer range for spin qubits with quantization axes along the $x$ direction for the device gate geometry, which is opposite to the thin film results. 
The above results indicate the importance of volume integral equation based methods to characterize correlations in low-frequency electromagnetic fluctuations in the nano-electromagnetic environment, and their importance in device engineering to suppress noise correlations detrimental to quantum error correction.

\section{Controlling Noise Effects in Spin Qubit Quantum Sensing}\label{section_qs}

\begin{figure}[!t]
    \centering
    \includegraphics[width=3in]{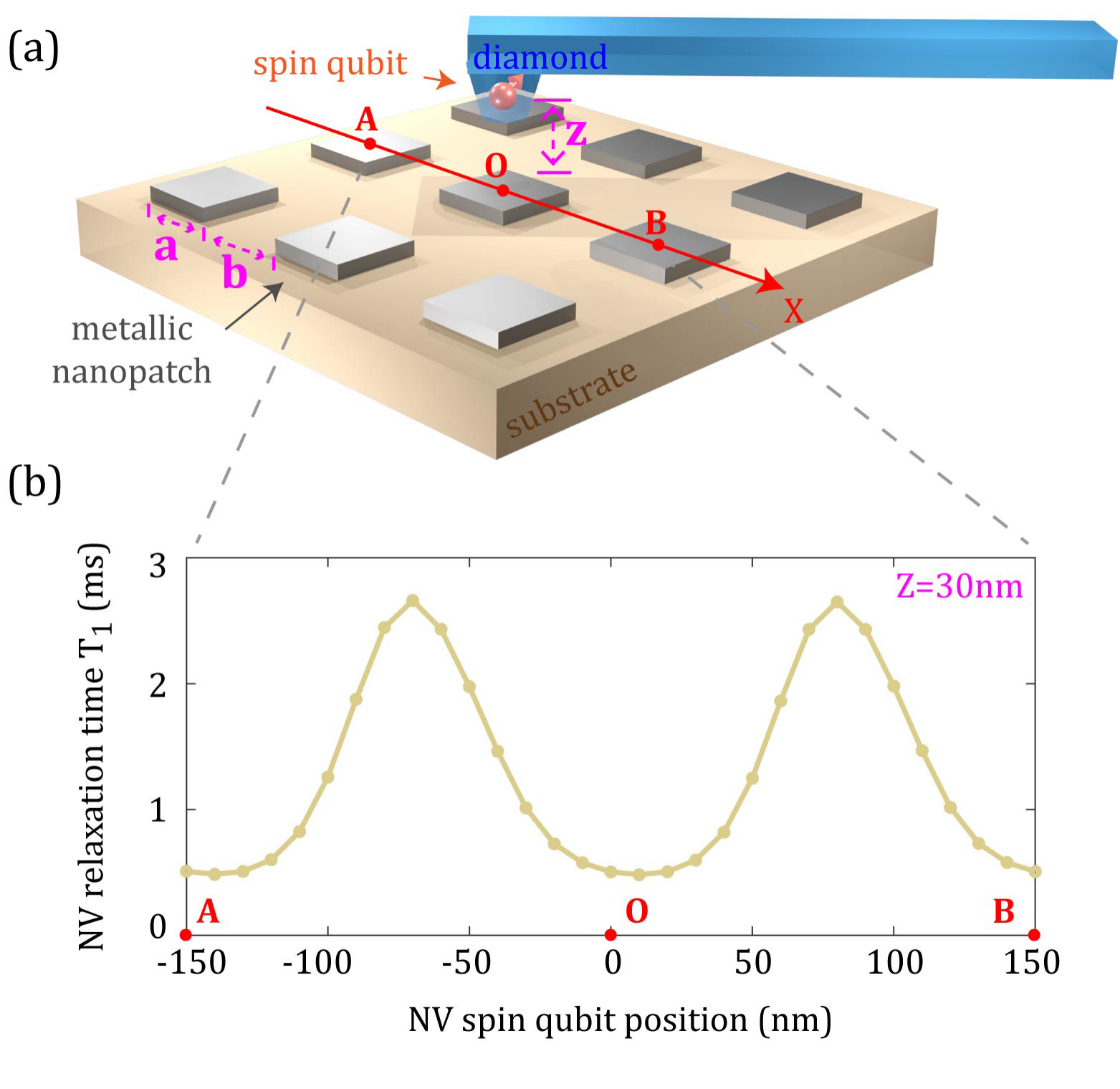}
    \caption{Modeling nanoscale conductivity imaging of nanopatterned metallic patches using NV centers. (a) A schematic of a single spin qubit at distance $z$ from nanopatterned metallic patches of size $a\times a$ with nearest neighbor distance $b$. (b) NV relaxation time $T_1$ simulated with volume integral equation based methods along the x-axis from point A to point B. Through the combination of experimental measurements and volume integral equation based simulations, the conductivity of nanostructures can be quantitatively probed by quantum sensors.}
    \label{fig:figure5}
\end{figure}

In this section, we demonstrate the application of our theoretical framework in controlling noise effects for quantum sensing applications. Different from quantum computing devices where noise needs to be suppressed, it is important to enhance the low-frequency magnetic noise for probing microscopic material properties in quantum sensing. 

In quantum relaxometry and dephasometry, material properties are probed by the differences in spin qubit relaxation and dephasing time in the near-field and far-field of the material. As a result, it is important to enhance the near-field low-frequency magnetic noise to ensure that the noise effects are large enough to exceed the sensitivity of spin qubits. Engineering material geometries in the nanoscale provides a possibility to enhance the noise effects associated with material properties of interest. Another important aspect of quantum sensing is its advantage in probing material properties at the nanoscale level. For example, nitrogen-vacancy (NV) center spin qubits have been widely used for nanoscale imaging and probing materials with strong inhomogeneity, such as superconductors with disconnected superconducting regions~\cite{bhattacharyya2024imaging}. However, the results predicted by Eq.~(\ref{anacpd}) under the thin film approximation are generally not valid in the presence of nanostructures or material inhomogeneity. Meanwhile, the volume integral equation based methods provide an approach to model and control low-frequency magnetic noise in nanoscale quantum sensing. 

To exemplify the application of our theoretical framework in quantitatively sensing material properties, we consider nanoscale conductivity imaging of nanopatterned metallic patches using NV centers~\cite{ariyaratne2018nanoscale}, which is important for measuring local conductivity of nanostructured materials without introducing any contact. As shown in Fig.~\ref{fig:figure5}(a), we consider NV spin qubits of frequencies around $2.87$GHz at distance $z$ from nanopatterned silver patches (grey) on top of bottom substrates. We consider metal patches of side length $a$ patterned with nearest neighbor distance $b$. We assume the metallic nanopatches to be polycrystalline and neglect nonlocality in their electromagnetic response. Meanwhile, in the case of single-crystal metal structures, nonlocal volume integral equation based solvers can be employed to capture nonlocal effects~\cite{zhou2023fast}. The conductivity of these metallic nanopatches can be imaged by measuring the relaxation time of NV spin qubits when they are scanned above the surface~\cite{ariyaratne2018nanoscale}.

In Fig.~\ref{fig:figure5}(b), we demonstrate that NV spin qubit relaxation time $T_1$ can be modeled by volume integral equations methods in this quantum sensing application. In this calculation, we employ 52100 tetrahedron elements, and the corresponding VIE system matrix takes around 185 GB of memory. The solution time using a 128-core workstation is around 1043 seconds for one spatial configuration of spin qubits. The NV relaxation time $T_1$ is related to relaxation rates by $T_1=[3(2\bar{\mathcal{N}}+1)\Gamma^{ii}_r]^{-1}$~\cite{kolkowitz2015probing}. Here, we consider $a=90\,\mathrm{nm}$, $b=60\,\mathrm{nm}$, and $z=30\,\mathrm{nm}$, and plot $T_1$ between points A and B separated by $300\,\mathrm{nm}$. As expected, we find large variations of $T_1$ when NV spin qubits are scanned along the x axis. Meanwhile, it is clear that this T1 behavior can not be modeled by the thin film approximation results. In this simulation, we consider silver conductivity $\sigma_{Ag}=5\times 10^7 \, \mathrm{S/m}$ at room temperature~\cite{de1988temperature}. This indicates that, with volume integral equation based simulations, the conductivity of nanostructures can be quantitatively probed by quantum sensors. The above results indicate the importance of employing volume integral equation based methods for quantitatively probing material properties.

\section{Conclusion}\label{section_conclusion}
In conclusion, we introduced a numerical framework to engineer low-frequency magnetic noise for enhancing spin qubit quantum sensing and computing performance by leveraging advanced computational electromagnetics techniques, especially fast and accurate volume integral equation based solvers. Our work extends the application of computational electromagnetics, especially volume integral equation based methods, to model noise effects in spin qubit quantum sensing and computing. We apply our numerical framework to model low-frequency magnetic noise in the nano-electromagnetic environment in realistic applications, including a semiconductor spin qubit quantum computing device and nanoscale imaging based on
quantum-impurity relaxometry. We demonstrate the limitation of the thin film approximation used in modeling magnetic noise in spin qubit systems, which can fail to predict qualitatively correct noise behaviors in realistic devices. Our work paves the way for device engineering to control magnetic fluctuation noise in spin qubit quantum applications. Beyond this, our numerical framework can also be integrated with topology optimization~\cite{jensen2011topology} to optimize the device design to benefit quantum sensing and computing.

\section*{Acknowledgment}
This work was supported by the Army Research Office under Grant No. W911NF-21-1-0287.

\ifCLASSOPTIONcaptionsoff
  \newpage
\fi



%

\bibliography{reference}

\begin{thebibliography}{10}
\providecommand{\url}[1]{#1}
\csname url@samestyle\endcsname
\providecommand{\newblock}{\relax}
\providecommand{\bibinfo}[2]{#2}
\providecommand{\BIBentrySTDinterwordspacing}{\spaceskip=0pt\relax}
\providecommand{\BIBentryALTinterwordstretchfactor}{4}
\providecommand{\BIBentryALTinterwordspacing}{\spaceskip=\fontdimen2\font plus
\BIBentryALTinterwordstretchfactor\fontdimen3\font minus \fontdimen4\font\relax}
\providecommand{\BIBforeignlanguage}[2]{{%
\expandafter\ifx\csname l@#1\endcsname\relax
\typeout{** WARNING: IEEEtran.bst: No hyphenation pattern has been}%
\typeout{** loaded for the language `#1'. Using the pattern for}%
\typeout{** the default language instead.}%
\else
\language=\csname l@#1\endcsname
\fi
#2}}
\providecommand{\BIBdecl}{\relax}
\BIBdecl

\bibitem{burkard2023semiconductor}
G.~Burkard, T.~D. Ladd, A.~Pan, J.~M. Nichol, and J.~R. Petta, ``Semiconductor spin qubits,'' \emph{Reviews of Modern Physics}, vol.~95, no.~2, p. 025003, 2023.

\bibitem{suter2016colloquium}
D.~Suter and G.~A. {\'A}lvarez, ``Colloquium: Protecting quantum information against environmental noise,'' \emph{Reviews of Modern Physics}, vol.~88, no.~4, p. 041001, 2016.

\bibitem{chekhovich2013nuclear}
E.~Chekhovich, M.~Makhonin, A.~Tartakovskii, A.~Yacoby, H.~Bluhm, K.~Nowack, and L.~Vandersypen, ``Nuclear spin effects in semiconductor quantum dots,'' \emph{Nature materials}, vol.~12, no.~6, pp. 494--504, 2013.

\bibitem{itoh2014isotope}
K.~M. Itoh and H.~Watanabe, ``Isotope engineering of silicon and diamond for quantum computing and sensing applications,'' \emph{MRS communications}, vol.~4, no.~4, pp. 143--157, 2014.

\bibitem{kuhlmann2013charge}
A.~V. Kuhlmann, J.~Houel, A.~Ludwig, L.~Greuter, D.~Reuter, A.~D. Wieck, M.~Poggio, and R.~J. Warburton, ``Charge noise and spin noise in a semiconductor quantum device,'' \emph{Nature Physics}, vol.~9, no.~9, pp. 570--575, 2013.

\bibitem{yoneda2023noise}
J.~Yoneda, J.~Rojas-Arias, P.~Stano, K.~Takeda, A.~Noiri, T.~Nakajima, D.~Loss, and S.~Tarucha, ``Noise-correlation spectrum for a pair of spin qubits in silicon,'' \emph{Nature Physics}, vol.~19, no.~12, pp. 1793--1798, 2023.

\bibitem{boter2020spatial}
J.~M. Boter, X.~Xue, T.~Kr{\"a}henmann, T.~F. Watson, V.~N. Premakumar, D.~R. Ward, D.~E. Savage, M.~G. Lagally, M.~Friesen, S.~N. Coppersmith \emph{et~al.}, ``Spatial noise correlations in a si/sige two-qubit device from bell state coherences,'' \emph{Physical Review B}, vol. 101, no.~23, p. 235133, 2020.

\bibitem{paquelet2023reducing}
B.~Paquelet~Wuetz, D.~Degli~Esposti, A.-M.~J. Zwerver, S.~V. Amitonov, M.~Botifoll, J.~Arbiol, A.~Sammak, L.~M. Vandersypen, M.~Russ, and G.~Scappucci, ``Reducing charge noise in quantum dots by using thin silicon quantum wells,'' \emph{Nature communications}, vol.~14, no.~1, p. 1385, 2023.

\bibitem{connors2019low}
E.~J. Connors, J.~Nelson, H.~Qiao, L.~F. Edge, and J.~M. Nichol, ``Low-frequency charge noise in si/sige quantum dots,'' \emph{Physical Review B}, vol. 100, no.~16, p. 165305, 2019.

\bibitem{langsjoen2012qubit}
L.~S. Langsjoen, A.~Poudel, M.~G. Vavilov, and R.~Joynt, ``Qubit relaxation from evanescent-wave johnson noise,'' \emph{Physical Review A}, vol.~86, no.~1, p. 010301, 2012.

\bibitem{sun2023limits}
W.~Sun, S.~Bharadwaj, L.-P. Yang, Y.-L. Hsueh, Y.~Wang, D.~Jiao, R.~Rahman, and Z.~Jacob, ``Limits to quantum gate fidelity from near-field thermal and vacuum fluctuations,'' \emph{Physical Review Applied}, vol.~19, no.~6, p. 064038, 2023.

\bibitem{degen2017quantum}
C.~L. Degen, F.~Reinhard, and P.~Cappellaro, ``Quantum sensing,'' \emph{Reviews of modern physics}, vol.~89, no.~3, p. 035002, 2017.

\bibitem{kolkowitz2015probing}
S.~Kolkowitz, A.~Safira, A.~High, R.~Devlin, S.~Choi, Q.~Unterreithmeier, D.~Patterson, A.~Zibrov, V.~Manucharyan, H.~Park \emph{et~al.}, ``Probing johnson noise and ballistic transport in normal metals with a single-spin qubit,'' \emph{Science}, vol. 347, no. 6226, pp. 1129--1132, 2015.

\bibitem{tenberg2019electron}
S.~B. Tenberg, S.~Asaad, M.~T. M{{a}}dzik, M.~A. Johnson, B.~Joecker, A.~Laucht, F.~E. Hudson, K.~M. Itoh, A.~M. Jakob, B.~C. Johnson \emph{et~al.}, ``Electron spin relaxation of single phosphorus donors in metal-oxide-semiconductor nanoscale devices,'' \emph{Physical Review B}, vol.~99, no.~20, p. 205306, 2019.

\bibitem{takeda2021quantum}
K.~Takeda, A.~Noiri, T.~Nakajima, J.~Yoneda, T.~Kobayashi, and S.~Tarucha, ``Quantum tomography of an entangled three-qubit state in silicon,'' \emph{Nature Nanotechnology}, vol.~16, no.~9, pp. 965--969, 2021.

\bibitem{huang2019fidelity}
W.~Huang, C.~Yang, K.~Chan, T.~Tanttu, B.~Hensen, R.~Leon, M.~Fogarty, J.~Hwang, F.~Hudson, K.~M. Itoh \emph{et~al.}, ``Fidelity benchmarks for two-qubit gates in silicon,'' \emph{Nature}, vol. 569, no. 7757, pp. 532--536, 2019.

\bibitem{he2019two}
Y.~He, S.~Gorman, D.~Keith, L.~Kranz, J.~Keizer, and M.~Simmons, ``A two-qubit gate between phosphorus donor electrons in silicon,'' \emph{Nature}, vol. 571, no. 7765, pp. 371--375, 2019.

\bibitem{morello2020donor}
A.~Morello, J.~J. Pla, P.~Bertet, and D.~N. Jamieson, ``Donor spins in silicon for quantum technologies,'' \emph{Advanced Quantum Technologies}, vol.~3, no.~11, p. 2000005, 2020.

\bibitem{wang2016highly}
Y.~Wang, A.~Tankasala, L.~C. Hollenberg, G.~Klimeck, M.~Y. Simmons, and R.~Rahman, ``Highly tunable exchange in donor qubits in silicon,'' \emph{npj Quantum Information}, vol.~2, no.~1, pp. 1--5, 2016.

\bibitem{xue2022quantum}
X.~Xue, M.~Russ, N.~Samkharadze, B.~Undseth, A.~Sammak, G.~Scappucci, and L.~M. Vandersypen, ``Quantum logic with spin qubits crossing the surface code threshold,'' \emph{Nature}, vol. 601, no. 7893, pp. 343--347, 2022.

\bibitem{hendrickx2021four}
N.~W. Hendrickx, W.~I. Lawrie, M.~Russ, F.~van Riggelen, S.~L. de~Snoo, R.~N. Schouten, A.~Sammak, G.~Scappucci, and M.~Veldhorst, ``A four-qubit germanium quantum processor,'' \emph{Nature}, vol. 591, no. 7851, pp. 580--585, 2021.

\bibitem{awschalom2021quantum}
D.~D. Awschalom, C.~R. Du, R.~He, F.~J. Heremans, A.~Hoffmann, J.~Hou, H.~Kurebayashi, Y.~Li, L.~Liu, V.~Novosad \emph{et~al.}, ``Quantum engineering with hybrid magnonic systems and materials,'' \emph{IEEE Transactions on Quantum Engineering}, vol.~2, pp. 1--36, 2021.

\bibitem{niknam2022quantum}
M.~Niknam, M.~F.~F. Chowdhury, M.~M. Rajib, W.~A. Misba, R.~N. Schwartz, K.~L. Wang, J.~Atulasimha, and L.-S. Bouchard, ``Quantum control of spin qubits using nanomagnets,'' \emph{Communications Physics}, vol.~5, no.~1, p. 284, 2022.

\bibitem{sun2024nano}
W.~Sun, A.~E.~R. L{\'o}pez, and Z.~Jacob, ``Nano-electromagnetic super-dephasing in collective atom-atom interactions,'' \emph{arXiv preprint arXiv:2402.18816}, 2024.

\bibitem{khosravi2024giant}
F.~Khosravi, W.~Sun, C.~Khandekar, T.~Li, and Z.~Jacob, ``Giant enhancement of vacuum friction in spinning yig nanospheres,'' \emph{New Journal of Physics}, vol.~26, no.~5, p. 053006, may 2024.

\bibitem{premakumar2017evanescent}
V.~N. Premakumar, M.~G. Vavilov, and R.~Joynt, ``Evanescent-wave johnson noise in small devices,'' \emph{Quantum Science and Technology}, vol.~3, no.~1, p. 015001, 2017.

\bibitem{machado2023quantum}
F.~Machado, E.~A. Demler, N.~Y. Yao, and S.~Chatterjee, ``Quantum noise spectroscopy of dynamical critical phenomena,'' \emph{Physical Review Letters}, vol. 131, no.~7, p. 070801, 2023.

\bibitem{dolgirev2023local}
P.~E. Dolgirev, I.~Esterlis, A.~A. Zibrov, M.~D. Lukin, T.~Giamarchi, and E.~Demler, ``Local noise spectroscopy of wigner crystals in two-dimensional materials,'' \emph{Phys. Rev. Lett.}, vol. 132, p. 246504, Jun 2024.

\bibitem{ariyaratne2018nanoscale}
A.~Ariyaratne, D.~Bluvstein, B.~A. Myers, and A.~C.~B. Jayich, ``Nanoscale electrical conductivity imaging using a nitrogen-vacancy center in diamond,'' \emph{Nature communications}, vol.~9, no.~1, p. 2406, 2018.

\bibitem{van2015nanometre}
T.~Van~der Sar, F.~Casola, R.~Walsworth, and A.~Yacoby, ``Nanometre-scale probing of spin waves using single electron spins,'' \emph{Nature communications}, vol.~6, no.~1, p. 7886, 2015.

\bibitem{dwyer2022probing}
B.~L. Dwyer, L.~V. Rodgers, E.~K. Urbach, D.~Bluvstein, S.~Sangtawesin, H.~Zhou, Y.~Nassab, M.~Fitzpatrick, Z.~Yuan, K.~De~Greve \emph{et~al.}, ``Probing spin dynamics on diamond surfaces using a single quantum sensor,'' \emph{PRX Quantum}, vol.~3, no.~4, p. 040328, 2022.

\bibitem{staudacher2015probing}
T.~Staudacher, N.~Raatz, S.~Pezzagna, J.~Meijer, F.~Reinhard, C.~Meriles, and J.~Wrachtrup, ``Probing molecular dynamics at the nanoscale via an individual paramagnetic centre,'' \emph{Nature communications}, vol.~6, no.~1, p. 8527, 2015.

\bibitem{bhattacharyya2024imaging}
P.~Bhattacharyya, W.~Chen, X.~Huang, S.~Chatterjee, B.~Huang, B.~Kobrin, Y.~Lyu, T.~Smart, M.~Block, E.~Wang \emph{et~al.}, ``Imaging the meissner effect in hydride superconductors using quantum sensors,'' \emph{Nature}, pp. 1--7, 2024.

\bibitem{thiel2019probing}
L.~Thiel, Z.~Wang, M.~A. Tschudin, D.~Rohner, I.~Guti{\'e}rrez-Lezama, N.~Ubrig, M.~Gibertini, E.~Giannini, A.~F. Morpurgo, and P.~Maletinsky, ``Probing magnetism in 2d materials at the nanoscale with single-spin microscopy,'' \emph{Science}, vol. 364, no. 6444, pp. 973--976, 2019.

\bibitem{rovny2024new}
J.~Rovny, S.~Gopalakrishnan, A.~C.~B. Jayich, P.~Maletinsky, E.~Demler, and N.~P. de~Leon, ``New opportunities in condensed matter physics for nanoscale quantum sensors,'' \emph{arXiv preprint arXiv:2403.13710}, 2024.

\bibitem{baranov2017modifying}
D.~G. Baranov, R.~S. Savelev, S.~V. Li, A.~E. Krasnok, and A.~Al{\`u}, ``Modifying magnetic dipole spontaneous emission with nanophotonic structures,'' \emph{Laser \& Photonics Reviews}, vol.~11, no.~3, p. 1600268, 2017.

\bibitem{boddeti2021long}
A.~K. Boddeti, J.~Guan, T.~Sentz, X.~Juarez, W.~Newman, C.~Cortes, T.~W. Odom, and Z.~Jacob, ``Long-range dipole--dipole interactions in a plasmonic lattice,'' \emph{Nano letters}, vol.~22, no.~1, pp. 22--28, 2021.

\bibitem{cortes2022fundamental}
C.~L. Cortes, W.~Sun, and Z.~Jacob, ``Fundamental efficiency bound for quantum coherent energy transfer in nanophotonics,'' \emph{Optics Express}, vol.~30, no.~19, pp. 34\,725--34\,739, 2022.

\bibitem{otey2014fluctuational}
C.~R. Otey, L.~Zhu, S.~Sandhu, and S.~Fan, ``Fluctuational electrodynamics calculations of near-field heat transfer in non-planar geometries: A brief overview,'' \emph{Journal of Quantitative Spectroscopy and Radiative Transfer}, vol. 132, pp. 3--11, 2014.

\bibitem{rodriguez2011frequency}
A.~W. Rodriguez, O.~Ilic, P.~Bermel, I.~Celanovic, J.~D. Joannopoulos, M.~Solja{\v{c}}i{\'c}, and S.~G. Johnson, ``Frequency-selective near-field radiative heat transfer between photonic crystal slabs: a computational approach for arbitrary geometries and materials,'' \emph{Physical review letters}, vol. 107, no.~11, p. 114302, 2011.

\bibitem{reid2009efficient}
M.~H. Reid, A.~W. Rodriguez, J.~White, and S.~G. Johnson, ``Efficient computation of casimir interactions between arbitrary 3d objects,'' \emph{Physical review letters}, vol. 103, no.~4, p. 040401, 2009.

\bibitem{roth2021full}
T.~E. Roth and W.~C. Chew, ``Full-wave computation of the spontaneous emission rate of a transmon qubit,'' in \emph{2021 IEEE International Symposium on Antennas and Propagation and USNC-URSI Radio Science Meeting (APS/URSI)}.\hskip 1em plus 0.5em minus 0.4em\relax IEEE, 2021, pp. 1801--1802.

\bibitem{pham2023spectral}
D.~N. Pham, R.~D. Li, and H.~E. T{\"u}reci, ``Spectral theory for non-linear superconducting microwave systems: Extracting relaxation rates and mode hybridization,'' \emph{arXiv preprint arXiv:2309.03435}, 2023.

\bibitem{wang2022towards}
C.~Wang, X.~Li, H.~Xu, Z.~Li, J.~Wang, Z.~Yang, Z.~Mi, X.~Liang, T.~Su, C.~Yang \emph{et~al.}, ``Towards practical quantum computers: Transmon qubit with a lifetime approaching 0.5 milliseconds,'' \emph{npj Quantum Information}, vol.~8, no.~1, p.~3, 2022.

\bibitem{2018_Maiomiao_Direct}
M.~Ma and D.~Jiao, ``Accuracy directly controlled fast direct solution of general $\mathcal{H}^{2}$ -matrices and its application to solving electrodynamic volume integral equations,'' \emph{IEEE Transactions on Microwave Theory and Techniques}, vol.~66, no.~1, pp. 35--48, 2018.

\bibitem{omar2015linear}
S.~Omar and D.~Jiao, ``A linear complexity direct volume integral equation solver for full-wave 3-d circuit extraction in inhomogeneous materials,'' \emph{IEEE Transactions on Microwave Theory and Techniques}, vol.~63, no.~3, pp. 897--912, 2015.

\bibitem{YifanWang_TAP2022}
Y.~Wang and D.~Jiao, ``Fast {O(N logN)} algorithm for generating rank-minimized $\mathrm{H}^2$-representation of electrically large volume integral equations,'' \emph{IEEE Transactions on Antennas and Propagation}, vol.~70, no.~8, pp. 6944--6956, 2022.

\bibitem{tosi2023antenna}
L.~Tosi and P.~Rocca, ``Antenna array analysis through universal quantum computing processors—a study on noise modeling and impact,'' \emph{IEEE Transactions on Microwave Theory and Techniques}, 2023.

\bibitem{solgun2014blackbox}
F.~Solgun, D.~W. Abraham, and D.~P. DiVincenzo, ``Blackbox quantization of superconducting circuits using exact impedance synthesis,'' \emph{Physical Review B}, vol.~90, no.~13, p. 134504, 2014.

\bibitem{smith2016quantization}
W.~Smith, A.~Kou, U.~Vool, I.~Pop, L.~Frunzio, R.~Schoelkopf, and M.~Devoret, ``Quantization of inductively shunted superconducting circuits,'' \emph{Physical Review B}, vol.~94, no.~14, p. 144507, 2016.

\bibitem{ryu2023matrix}
C.~J. Ryu, D.-Y. Na, and W.~C. Chew, ``Matrix product states and numerical mode decomposition for the analysis of gauge-invariant cavity quantum electrodynamics,'' \emph{Physical Review A}, vol. 107, no.~6, p. 063707, 2023.

\bibitem{na2021diagonalization}
D.-Y. Na, J.~Zhu, and W.~C. Chew, ``Diagonalization of the hamiltonian for finite-sized dispersive media: Canonical quantization with numerical mode decomposition,'' \emph{Physical Review A}, vol. 103, no.~6, p. 063707, 2021.

\bibitem{daley2022practical}
A.~J. Daley, I.~Bloch, C.~Kokail, S.~Flannigan, N.~Pearson, M.~Troyer, and P.~Zoller, ``Practical quantum advantage in quantum simulation,'' \emph{Nature}, vol. 607, no. 7920, pp. 667--676, 2022.

\bibitem{bauer2023quantum}
C.~W. Bauer, Z.~Davoudi, A.~B. Balantekin, T.~Bhattacharya, M.~Carena, W.~A. De~Jong, P.~Draper, A.~El-Khadra, N.~Gemelke, M.~Hanada \emph{et~al.}, ``Quantum simulation for high-energy physics,'' \emph{PRX quantum}, vol.~4, no.~2, p. 027001, 2023.

\bibitem{bauer2020quantum}
B.~Bauer, S.~Bravyi, M.~Motta, and G.~K.-L. Chan, ``Quantum algorithms for quantum chemistry and quantum materials science,'' \emph{Chemical Reviews}, vol. 120, no.~22, pp. 12\,685--12\,717, 2020.

\bibitem{von2021quantum}
V.~von Burg, G.~H. Low, T.~H{\"a}ner, D.~S. Steiger, M.~Reiher, M.~Roetteler, and M.~Troyer, ``Quantum computing enhanced computational catalysis,'' \emph{Physical Review Research}, vol.~3, no.~3, p. 033055, 2021.

\bibitem{delgado2022simulating}
A.~Delgado, P.~A. Casares, R.~Dos~Reis, M.~S. Zini, R.~Campos, N.~Cruz-Hern{\'a}ndez, A.-C. Voigt, A.~Lowe, S.~Jahangiri, M.~A. Martin-Delgado \emph{et~al.}, ``Simulating key properties of lithium-ion batteries with a fault-tolerant quantum computer,'' \emph{Physical Review A}, vol. 106, no.~3, p. 032428, 2022.

\bibitem{rosenberg2020solid}
D.~Rosenberg, S.~J. Weber, D.~Conway, D.-R.~W. Yost, J.~Mallek, G.~Calusine, R.~Das, D.~Kim, M.~E. Schwartz, W.~Woods \emph{et~al.}, ``Solid-state qubits: 3d integration and packaging,'' \emph{IEEE Microwave Magazine}, vol.~21, no.~8, pp. 72--85, 2020.

\bibitem{chatterjee2021semiconductor}
A.~Chatterjee, P.~Stevenson, S.~De~Franceschi, A.~Morello, N.~P. de~Leon, and F.~Kuemmeth, ``Semiconductor qubits in practice,'' \emph{Nature Reviews Physics}, vol.~3, no.~3, pp. 157--177, 2021.

\bibitem{zwerver2022qubits}
A.~Zwerver, T.~Kr{\"a}henmann, T.~Watson, L.~Lampert, H.~C. George, R.~Pillarisetty, S.~Bojarski, P.~Amin, S.~Amitonov, J.~Boter \emph{et~al.}, ``Qubits made by advanced semiconductor manufacturing,'' \emph{Nature Electronics}, vol.~5, no.~3, pp. 184--190, 2022.

\bibitem{neyens2024probing}
S.~Neyens, O.~K. Zietz, T.~F. Watson, F.~Luthi, A.~Nethwewala, H.~C. George, E.~Henry, M.~Islam, A.~J. Wagner, F.~Borjans \emph{et~al.}, ``Probing single electrons across 300-mm spin qubit wafers,'' \emph{Nature}, vol. 629, no. 8010, pp. 80--85, 2024.

\bibitem{sun2024full}
B.~Sun, T.~Brecht, B.~H. Fong, M.~Akmal, J.~Z. Blumoff, T.~A. Cain, F.~W. Carter, D.~H. Finestone, M.~N. Fireman, W.~Ha \emph{et~al.}, ``Full-permutation dynamical decoupling in triple-quantum-dot spin qubits,'' \emph{PRX Quantum}, vol.~5, no.~2, p. 020356, 2024.

\bibitem{loss1998quantum}
D.~Loss and D.~P. DiVincenzo, ``Quantum computation with quantum dots,'' \emph{Physical Review A}, vol.~57, no.~1, p. 120, 1998.

\bibitem{kane1998silicon}
B.~E. Kane, ``A silicon-based nuclear spin quantum computer,'' \emph{Nature}, vol. 393, no. 6681, pp. 133--137, 1998.

\bibitem{petta2005coherent}
J.~R. Petta, A.~C. Johnson, J.~M. Taylor, E.~A. Laird, A.~Yacoby, M.~D. Lukin, C.~M. Marcus, M.~P. Hanson, and A.~C. Gossard, ``Coherent manipulation of coupled electron spins in semiconductor quantum dots,'' \emph{Science}, vol. 309, no. 5744, pp. 2180--2184, 2005.

\bibitem{bacon2000universal}
D.~Bacon, J.~Kempe, D.~A. Lidar, and K.~B. Whaley, ``Universal fault-tolerant quantum computation on decoherence-free subspaces,'' \emph{Physical Review Letters}, vol.~85, no.~8, p. 1758, 2000.

\bibitem{miao2020universal}
K.~C. Miao, J.~P. Blanton, C.~P. Anderson, A.~Bourassa, A.~L. Crook, G.~Wolfowicz, H.~Abe, T.~Ohshima, and D.~D. Awschalom, ``Universal coherence protection in a solid-state spin qubit,'' \emph{Science}, vol. 369, no. 6510, pp. 1493--1497, 2020.

\bibitem{gali2019ab}
{\'A}.~Gali, ``Ab initio theory of the nitrogen-vacancy center in diamond,'' \emph{Nanophotonics}, vol.~8, no.~11, pp. 1907--1943, 2019.

\bibitem{childress2013diamond}
L.~Childress and R.~Hanson, ``Diamond nv centers for quantum computing and quantum networks,'' \emph{MRS bulletin}, vol.~38, no.~2, pp. 134--138, 2013.

\bibitem{herbschleb2019ultra}
E.~Herbschleb, H.~Kato, Y.~Maruyama, T.~Danjo, T.~Makino, S.~Yamasaki, I.~Ohki, K.~Hayashi, H.~Morishita, M.~Fujiwara \emph{et~al.}, ``Ultra-long coherence times amongst room-temperature solid-state spins,'' \emph{Nature communications}, vol.~10, no.~1, p. 3766, 2019.

\bibitem{debroux2021quantum}
R.~Debroux, C.~P. Michaels, C.~M. Purser, N.~Wan, M.~E. Trusheim, J.~A. Mart{\'\i}nez, R.~A. Parker, A.~M. Stramma, K.~C. Chen, L.~De~Santis \emph{et~al.}, ``Quantum control of the tin-vacancy spin qubit in diamond,'' \emph{Physical Review X}, vol.~11, no.~4, p. 041041, 2021.

\bibitem{anderson2022five}
C.~P. Anderson, E.~O. Glen, C.~Zeledon, A.~Bourassa, Y.~Jin, Y.~Zhu, C.~Vorwerk, A.~L. Crook, H.~Abe, J.~Ul-Hassan \emph{et~al.}, ``Five-second coherence of a single spin with single-shot readout in silicon carbide,'' \emph{Science advances}, vol.~8, no.~5, p. eabm5912, 2022.

\bibitem{vaidya2023quantum}
S.~Vaidya, X.~Gao, S.~Dikshit, I.~Aharonovich, and T.~Li, ``Quantum sensing and imaging with spin defects in hexagonal boron nitride,'' \emph{Advances in Physics: X}, vol.~8, no.~1, p. 2206049, 2023.

\bibitem{klesse2005quantum}
R.~Klesse and S.~Frank, ``Quantum error correction in spatially correlated quantum noise,'' \emph{Physical review letters}, vol.~95, no.~23, p. 230503, 2005.

\bibitem{buhmann2007dispersion}
S.~Y. Buhmann and D.-G. Welsch, ``Dispersion forces in macroscopic quantum electrodynamics,'' \emph{Progress in quantum electronics}, vol.~31, no.~2, pp. 51--130, 2007.

\bibitem{clerk2020hybrid}
A.~Clerk, K.~Lehnert, P.~Bertet, J.~Petta, and Y.~Nakamura, ``Hybrid quantum systems with circuit quantum electrodynamics,'' \emph{Nature Physics}, vol.~16, no.~3, pp. 257--267, 2020.

\bibitem{van2020electromagnetic}
T.~Van Der~Sijs, O.~El~Gawhary, and H.~Urbach, ``Electromagnetic scattering beyond the weak regime: Solving the problem of divergent born perturbation series by pad{\'e} approximants,'' \emph{Physical Review Research}, vol.~2, no.~1, p. 013308, 2020.

\bibitem{kleinman1990convergent}
R.~Kleinman, G.~Roach, and P.~Van Den~Berg, ``Convergent born series for large refractive indices,'' \emph{JOSA A}, vol.~7, no.~5, pp. 890--897, 1990.

\bibitem{breuer2002theory}
H.-P. Breuer and F.~Petruccione, \emph{The theory of open quantum systems}.\hskip 1em plus 0.5em minus 0.4em\relax Oxford University Press, USA, 2002.

\bibitem{buhmann2012macroscopic}
S.~Y. Buhmann, D.~T. Butcher, and S.~Scheel, ``Macroscopic quantum electrodynamics in nonlocal and nonreciprocal media,'' \emph{New Journal of Physics}, vol.~14, no.~8, p. 083034, 2012.

\bibitem{scheel2009macroscopic}
S.~Scheel and S.~Y. Buhmann, ``Macroscopic qed-concepts and applications,'' \emph{arXiv preprint arXiv:0902.3586}, 2009.

\bibitem{yang2020single}
L.-P. Yang, C.~Khandekar, T.~Li, and Z.~Jacob, ``Single photon pulse induced transient entanglement force,'' \emph{New Journal of Physics}, vol.~22, no.~2, p. 023037, 2020.

\bibitem{nielsen2002simple}
M.~A. Nielsen, ``A simple formula for the average gate fidelity of a quantum dynamical operation,'' \emph{Physics Letters A}, vol. 303, no.~4, pp. 249--252, 2002.

\bibitem{poudel2013relaxation}
A.~Poudel, L.~S. Langsjoen, M.~G. Vavilov, and R.~Joynt, ``Relaxation in quantum dots due to evanescent-wave johnson noise,'' \emph{Physical Review B}, vol.~87, no.~4, p. 045301, 2013.

\bibitem{khandekar2019thermal}
C.~Khandekar and Z.~Jacob, ``Thermal spin photonics in the near-field of nonreciprocal media,'' \emph{New Journal of Physics}, vol.~21, no.~10, p. 103030, 2019.

\bibitem{de1988temperature}
J.~De~Vries, ``Temperature and thickness dependence of the resistivity of thin polycrystalline aluminium, cobalt, nickel, palladium, silver and gold films,'' \emph{Thin Solid Films}, vol. 167, no. 1-2, pp. 25--32, 1988.

\bibitem{wood2018quantification}
C.~J. Wood and J.~M. Gambetta, ``Quantification and characterization of leakage errors,'' \emph{Physical Review A}, vol.~97, no.~3, p. 032306, 2018.

\bibitem{zhou2023fast}
R.~Zhou, W.~Sun, S.~Bharadwaj, Z.~Jacob, and D.~Jiao, ``Fast volume integral equation based modeling of quantum gate circuitry: Capturing local vs. nonlocal effects on spin qubits,'' in \emph{2023 IEEE International Symposium on Antennas and Propagation and USNC-URSI Radio Science Meeting (USNC-URSI)}.\hskip 1em plus 0.5em minus 0.4em\relax IEEE, 2023, pp. 1179--1180.

\bibitem{jensen2011topology}
J.~S. Jensen and O.~Sigmund, ``Topology optimization for nano-photonics,'' \emph{Laser \& Photonics Reviews}, vol.~5, no.~2, pp. 308--321, 2011.

\end{thebibliography}
\bibliographystyle{IEEEtran}




\end{document}